\documentclass[usegraphicx,usenatbib]{mn2e}    

\begin{document}

\newcommand{\lele}[3]{{#1}\,$\le$\,{#2}\,$\le$\,{#3}}
\newcommand{\tauav}[1]{$\tau_{#1}$/A$_{\rm V}$}
\newcommand{\iav}[1]{I$_{#1}$/A$_{\rm V}$}

%\headnote{Letter}
\title{Constraints on the nature of dust particles by infrared observations}

 % \thanks{Based on observations with ISO, an ESA project with instruments
 %   funded by ESA Member States (especially the PI countries:
 %   Germany, the Netherlands and the United Kingdom) and
 %   with the participation of ISAS and NASA.}  
%\subtitle{This is just a working title...}
%\subtitle{One should find a better title for this...}
%\author{Cs.~Kiss$^1$, P.~\'Abrah\'am$^1$, R.J.~Laureijs$^2$, S.M.~Birkmann$^3$}
  % \and Z.~Kiss\inst{3}
%\authorrunning{Cs. Kiss et al.}
%\titlerunning{Constraints on the nature of dust particles
%by infrared observations}
%\institute{ Konkoly Observatory of the Hungarian Academy of Sciences, 
%    P.O. Box 67, H-1525~Budapest, Hungary 
%    \and
%  European Space Agency, Astrophysics Division, Kepleraan 1, 
%    2201AZ Noordwijk, The Netherlands 
%    \and
%  Max-Planck-Institut f\"ur Astronomie, K\"onigstuhl 17, D-69117, 
%    	Heidelberg, Germany }
 % \and Astronomy Department, E\"otv\"os Lor\'and University, 
 %   P.O. Box 32, H-1518~Budapest, Hungary  
%\offprints{Cs.~Kiss, pkisscs@mpia.de}
%
%\date{ Received  / Accepted ...}
%
\author[Cs. Kiss et al.]
{Cs.~Kiss$^1$, P.~\'Abrah\'am$^1$, R.J.~Laureijs$^2$, A.~Mo\'or$^1$, 
  S.M.~Birkmann$^3$ 
\newauthor
\\
  $^1$Konkoly Observatory of the Hungarian Academy of Sciences, 
    P.O. Box 67, H-1525~Budapest, Hungary \\
  $^2$European Space Agency, Astrophysics Division, Keplerlaan 1, 
    2201AZ Noordwijk, The Netherlands \\
  $^3$Max-Planck-Institut f\"ur Astronomie, K\"onigstuhl 17, D-69117, 
    	Heidelberg, Germany }

\date{Accepted: 2006 September 24; Recieved: 2006 September 23;
 in original form: 2006 April 6}
%%%%%%%%%%%%%%%%%%%%%%%%%%%%%%%%%%%%%%%%%%%%%%%%%%%%%%%%%%%%%%%%%%

\maketitle
\begin{abstract}
The far-infrared (FIR) emissivity of dust is an important parameter
characterizing the physical properties of the grains. With the
availability of stellar databases and far-infrared data from 
Infrared Space Observatory (ISO) it is
possible to compare the optical and infrared properties of dust, and
derive the far-infrared emissivity with respect to the optical
extinction.
In this paper we present the results of a systematic analysis of the
FIR emissivity of interstellar clouds observed  with ISOPHOT (the
photometer onboard ISO) at least at two infrared wavelengths, one close
to $\sim$100$\mu$m and one at 200$\mu$m. We constructed FIR emission
maps, determined dust temperatures, created extinction maps using 2MASS 
survey data, and calculated far-infrared emissivity for each of these clouds. 
We present the largest homogeneously reduced database constructed so far for
this purpose. During the data analysis special care was taken on
possible systematic errors. We find that far-infrared emissivity 
has a clear dependence on temperature. 
The emissivity is enhanced by a factor of usually less than 2 in the low dust
temperature regime of 12K$\le$T$_{\rm d}$$\le$14K. This result suggests larger
grain sizes in those regions. However, the emissivity increase of 
typically below
2 restricts the possible grain growth processes to ice-mantle formation
and coagulation of silicate grains, and excludes the coagulation of
carbonaceous particles on the scales of the regions we investigated.
In the temperature range \lele{14\,K}{T$_{\rm d}$}{16\,K} a  systematic
decrease of emissivity is observed with respect to the  values of the
diffuse interstellar matter. Possible scenarios for this behaviour are
discussed in the paper. 
%\keywords{methods:\ observational  -- ISM:\ structure -- 
%          Infrared:\ ISM:\ continuum -- diffuse radiation}
\end{abstract}
%%%%%%%%%%%%%%%%%%%%%%%%%%%%%%%%%%%%%%%%%%%%%%%%%%%%%%%%%%%%%%%%%%

\section{Introduction}

%Dust emissivity

The far-infrared (FIR) emissivity of interstellar grains predominantly
emitting at wavelengths in excess of 100\,$\mu$m can now be determined
in many molecular and moderate density regions thanks to the increasing
availability of far-infrared, sub-mm, and massive stellar data sets.
Several studies have reported on the detection of an enhancement of the FIR
emissivity in the 100-200\,$\mu$m wavelength range in regions of higher column
density compared to the dust emissivity in the diffuse interstellar medium
associated with HI (Bernard et al., 1999; Cambr\'esy et al., 2001; Juvela et
al., 2002; Stepnik et al., 2003; Lehtinen et al., 2004, Rawlings et al.,
2005).
The largest sample of different regions is presented in del Burgo et
al. (2003) presenting a study of eight translucent clouds.

The observations suggest a trend where the emissivity increases
with decreasing temperature. 
The variation is attributed to a change in grain properties
which is expected to take place in denser environments.
In particular, the increase in emissivity is interpreted as
a signature of an enhancement in grain size.

However, the trend shows a large scatter which might be due to the
observational limitations which increase the uncertainties.
%
% note: there may be a scatter due to the nature of the
% sources and regions investigated.
%
On one hand, star counts statistically probe only a limited extinction
range and, on the other hand, the determination of the grain temperatures
in the infrared can have large uncertainties.
%
% The variation was inferred from the comparison
% of the ratio between far-infrared or submm opacity
% and visual extinction in different regions.
%
% Although simple two-compenent models predict a continous
% increase of emissivity with the decreasing temperature,
% observations show only a modest increase of $\epsilon$
% for low ($\sim$12K) temperatures. 
%

In this study we present a large sample based on 
Infrared Space Observatory (ISO) data of
cloud regions with reliable values of the FIR dust emissivity and optical
extinction data.  We examine in detail the possible observational and
data processing errors to ensure the reliability of our results. 
A general good agreement was found between our results and previous 
studies of individual regions. Due to
the large number of data points we are able to put stronger
constraints on the emissivity changes with temperature. 

%%%%%%%%%%%%%%%%%%%%%%%%%%%%%%%%%%%%%%%%%%%%%%%%%%%%%%%%%%%%%%%%%%%%%%%%%%%
  
\section{Observations and data reduction}

\subsection{Far-infrared maps}

We searched the ISO Archive (Salama 2004) for ISOPHOT observations
(Lemke et al. 1996)
of interstellar clouds matching the following criteria: (1) the field has
been covered at least at two far-infrared wavelengths: 
one at 200\,$\mu$m ({\it long}) and another 
either at 90, 100 or 120\,$\mu$m ({\it short})
in order to provide a sufficient wavelength interval for a reliable
colour temperature calculation; (2) the cloud has to be galactic and must
have sufficient dynamic range in brightness for correlation analyses; (3)
the map is larger than 5\arcmin~ at least in one dimension; and 
(4) there is no high mass star formation going on in
the vicinity of the cloud which could significantly change the local
interstellar radiation field. In total we selected 22 maps which is the
largest sample evaluated so far for studying far-infrared dust
emissivity. All selected maps were obtained with the 
P22 astronomical observing template mode (Laureijs et al.~2003). 
Measurement wavelengths, ISO-id numbers and central positions 
are listed in Table~1.

The ISOPHOT observations were performed with the C100
(43\farcs5$\times$43\farcs5 sized pixels) and C200 
(89\farcs5$\times$89\farcs5 sized pixels) cameras. 
The ISOPHOT data were processed with the Phot
Interactive Analysis software version 10.0 (PIA, Gabriel et al., 1997),
using standard batch processing and a first quartile flatfielding. We
followed in detail the processing scheme described in del Burgo et al.
(2003). The data were colour corrected taking into account the dust
temperature derived from the brightness ratio at the two wavelengths
(see Sect.~3.1). 

The officially quoted
absolute photometric uncertainty of the surface brightness in the
ISO Legacy Archive is $\le$20--25\% (Klaas et al., 2003). 
In the present study we estimated
independently this uncertainty in two ways. 
We compared the ISOPHOT and COBE/DIRBE background surface brightness 
values, interpolated to the ISOPHOT wavelengths (1) for a large sample 
of mini-map observations and (2) for our target fields. This analysis is presented 
in Appendix~A. The results of this investigation show, that the 
typical relative deviations of the individual ISOPHOT measurements
with respect to the COBE/DIRBE values is $\sim$15\%, and that there is no
noticable systematic discrepancy between the two photometric systems. 
Throughout this paper we present our results in the ISOPHOT/PIA\,10.0 
surface brightness photometric system. 
The effect of a potential imperfect surface brightness calibration 
is discussed in detail in Sect.~4.5, based on results presented 
in Appendix~A. 

\subsection{Extinction Data}

The traditional method to derive extinction in interstellar 
clouds is based on variations in stellar density 
in the sky due to the obscuration of dust (Wolf 1923). 
Star counts obtained by placing a regular grid on the target field 
are converted to B or V-band extinction using statistical methods; 
the zero extinction level is
obtained by comparison with a nearby extinction free reference field.
   
%Cambr\'esy et al. (1997) replaced the classical regular grid 
%by an adaptive one, which overcomes the limitation placed by the 
%too low count of stars in high column density regions. 
%However, this method can give only average extinction values at low spatial 
%resolution at the densest parts of the clouds. 
Cambr\'esy et al. (1997) replaced the classical regular grid 
with an adaptive one; in this method the gridsize is adjusted
to include a fixed number of stars, therefore it can still provide
extinction estimates for high density regions.  
However, in these cases the derived A$_{\rm V}$ values are averages
over the enlarged area of the adaptive cell. 

Due to the requirement of sufficient count of stars for 
reliable statistics, the minimum
resolution is limited to arcminute scales, and low dust column densities. The
maximum extinction in the visual is at best 5\,mag and cannot be improved
significantly by deep dedicated observations. In that respect, online
catalogues with optical stellar data are well suited for the method.

In recent years the application of near-infrared reddening of individual 
stars has become widely used due to the general availability of J, H
and K-band measurements. Extinction mapping methods using NIR reddening 
often combine the individual reddening values in some statistical manner in
order to minimize the effect of individual line of sights. 
The NICE and NICER colour excess methods (Lada et al. 1994, Lombardi \& Alves 2001)
proved to produce good quality maps in many applications. 
Although NIR star counts are more reliable for high extinction values, 
in the low to intermediate extinction range (A$_{\rm V}$\,$\le$\,15\,mag) the NIR colour
excess methods are superior over NIR star counts, since they are less affected
by the presence of foreground stars (Cambr\'esy et al. 2002). 

The 2MASS Point Source Catalogue (Cutri et al. 2003) is a powerful source of 
NIR data and can be used to obtain NIR colour excess / extinction as was done
by many authors in recent years.  

A detailed comparison of visual extinctions obtained by optical star count
and NIR colour excess is presented in Kiss et al. 
(2006)\footnote{Note, that the first author of the cited paper 
is \emph{not} the first author of the present paper}. 
In this paper extinction 
maps are derived using star counts of USNO data (Monet et al., 1998, 2003)
and using near-infrared (J, H and K$_{\rm S}$-band) reddening of 
2MASS data. 
Their fig.~2 shows a good correlation of the two datasets and the slope
of the scatter plot is very near to unity for low extinctions. 
A flattening of the USNO extinction occurs at 3...3.5\,mag, due 
to the already insufficient count statistics at these extinction
levels. 

The near-infrared colour excess methods are less affected by
uncertainties due to foreground stars and give reliable extinction values 
in denser regions up to $\sim$15\,mag. Therefore we decided to apply the NICER 
method on the 2MASS Point Source Catalogue to obtain extinction maps in this
paper. Following the findings by Cambr\'esy et al. (2002) we set 
A$_{\rm V}^{\rm lim}$\,=\,8\,mag for the liming magnitude of 2MASS data, assuming a 
10\% of foreground stars in the target field. Above this limit we did not
consider the extinction data to be reliable. These data points were
excluded from the further analysis.  Due to the low number of data points
above this limit and outlier resistant routines we used in the 
scatter plot analysis, difference between the effect of a cut in A$_{\rm V}$ or 
of a cut perpendicular to the slope at A$_{\rm V}^{\rm lim}$\,=\,8\,mag is negligible.

%%%%%%%%%%%%%%%%
\section{Results}

\subsection{Dust temperature}
%%
%%%%%%%%%%%
\begin{figure}
\includegraphics[width=8.5cm]{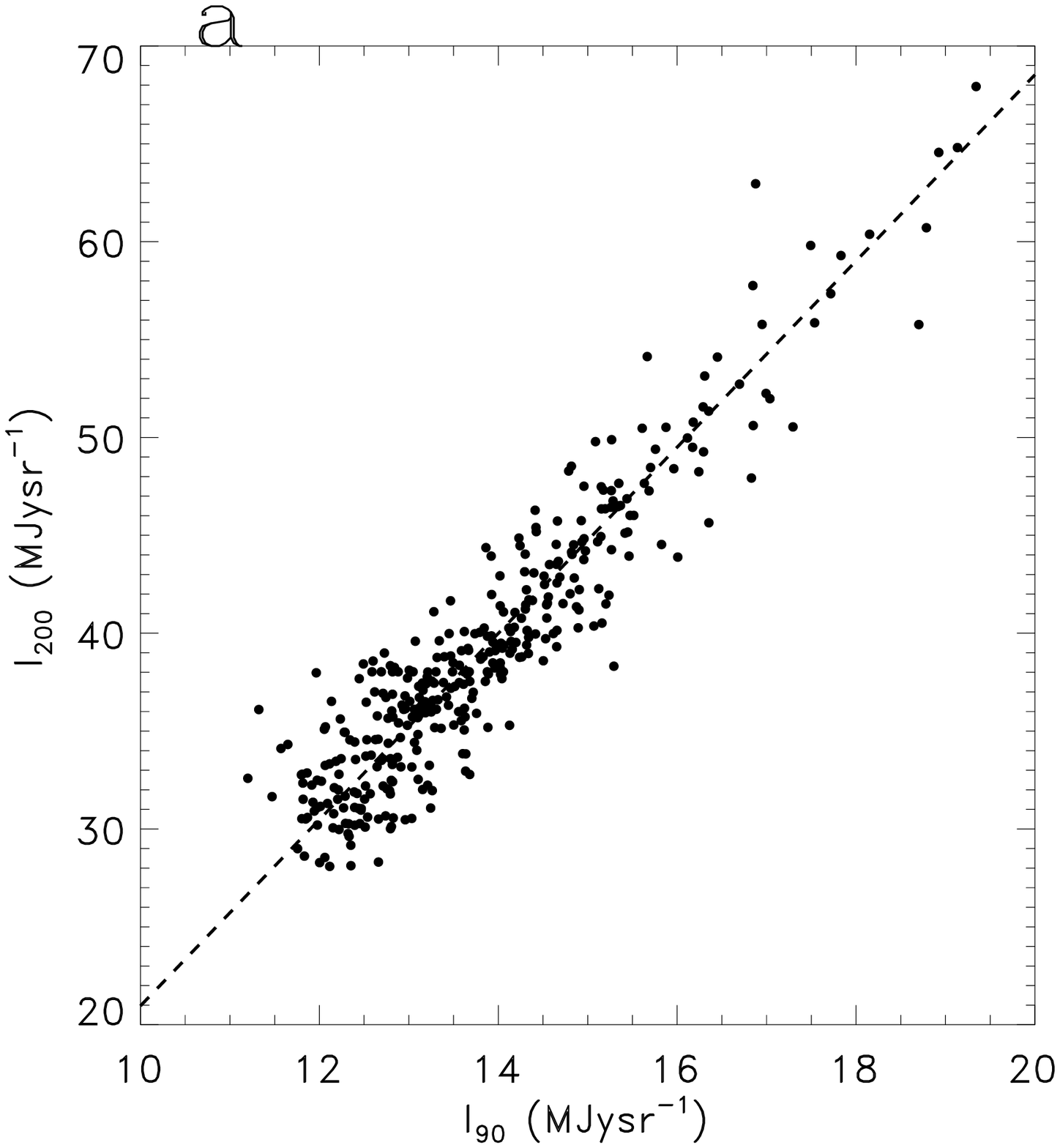}
\includegraphics[width=8.5cm]{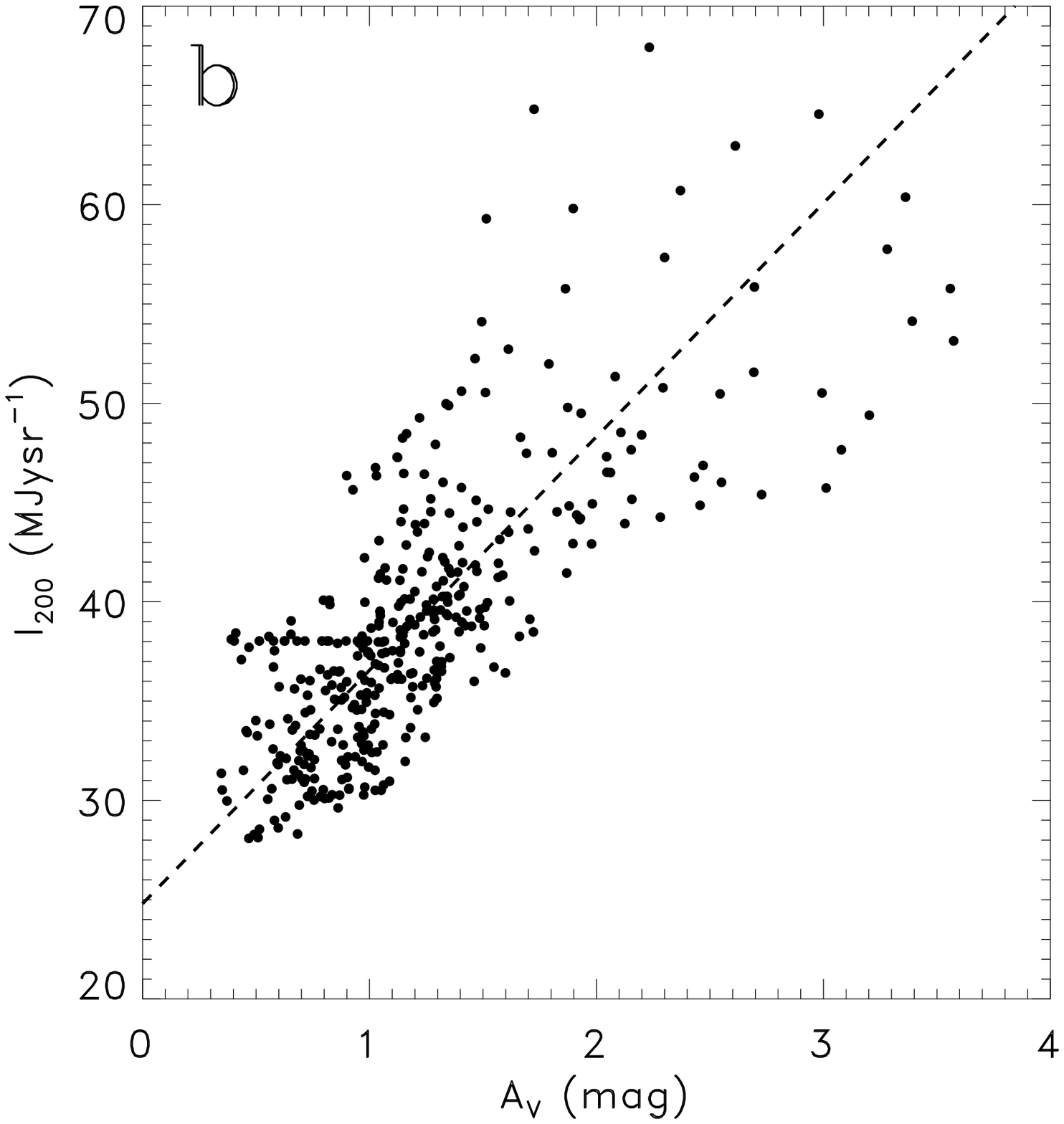}
\caption[]{Example ({\bf a}) of a short versus long wavelength 
surface brightness scatter plot and ({\bf b})
a visual extinction versus 200\,$\mu$m scatter plot, 
both in the G100.0+14.8 field. On both panels the dashed lines 
show the fitted linears.}
\label{fig:fit}
\end{figure}

We derived colour temperatures T$_{\rm d}$ from the slope in the
I$_{\lambda_1}$ versus I$_{\lambda_2}$ scatter plot of a specific
sky region, where I$_{\lambda_1}$ and I$_{\lambda_2}$ are the surface 
brightness values at the wavelengths $\lambda_1$ and $\lambda_2$,
respectively. An example in shown in Fig.~1a.
Maps taken with the C100 camera (90 and 100\,$\mu$m)
were smoothed to the resolution of the 200\,$\mu$m observation. 
The method of using scatter plots to determine T$_{\rm d}$ is described
in del Burgo et al. (2003). It has the main advantage that
the slope is insensitive to surface brightness offsets due to e.g.
zodiacal light, extragalactic background, calibrational zero
point, etc. 
The conversion of the $\Delta\rm I_{\lambda_1}/\Delta\rm I_{200}$
slopes to T$_{\rm d}$ was done via high resolution
$\Delta\rm I_{\lambda_1}/\Delta\rm I_{200}$-- T$_{\rm d}$ tables,
assuming a $\nu^{\beta}\rm B_{\nu}\rm(T)$ spectral energy distribution.
The tables also accounted for the correct colour correction. 
We adopted $\beta$\,=\,2.

In Sect.~4.3 we check the effect of a different emissivity law.
In the following we adopt T$_{\rm d}$ as an approximation of the physical
temperature of the dust grains. 

Del Burgo et al. (2003) found that the scatter plots of some fields can consist
of more than one linear section, indicating multiple dust temperatures within
the field. Inspection of the maps of these fields indicated that the different
temperature components came from interlocked regions which could be separated by A$_{\rm V}$
isocontours. The resulting I$_{\lambda_1}$ versus I$_{200}$ scatter plots of
the subfields were sufficiently linear. 

The uncertainty in the temperature T$_{\rm d}$ can be attributed to, firstly, the
error in the slope $\delta (\Delta\rm I_{\lambda_1} / \Delta\rm I_{200})$, and,
secondly, the calibration uncertainty in the surface brightness values.
The first component was computed from the formal uncertainty of the linear fits.
The second component was evaluated by using surface brightness uncertainty
values given in  Sect.~2.1. The resulting temperature and uncertainty values
are typically in the order of 0.7\,K depending on the colour temperature
(Table~1.)

\subsection{Emissivity parameters}
In Table~1 we list the derived values of
I$_{200}$/A$_{\rm V}$ obtained from the slopes of the I$_{200}$ versus A$_{\rm V}$
relationships in those regions where T$_{\rm d}$ is constant as delineated by a
constant I$_{\lambda_1}$ vs. I$_{200}$. The 200$\mu$m surface brightness
was smoothed to the resolution of A$_{\rm V}$. The ratio I$_{200}$/A$_{\rm V}$
consists of two independent observables and therefore also serves as a
good diagnostics of the main uncertainties.
An example A$_{\rm V}$ vs. I$_{200}$ scatter plot is presented in Fig.~1b.

 Using the measured value of T$_{\rm d}$ from the ratio $I_{\lambda_1}/I_{200}$,
we derive the ratio between the infrared and optical opacity
$\tau_{200}$/A$_{\rm V}$ from  $\tau_{200}$/A$_{\rm V}$\, =\,(I$_{200}$/A$_{\rm V}$)
$\times$ B$_{200}$(T$_{\rm d}$)$^{-1}$, where B$_{200}$(T$_{\rm d}$) is the Planck
function at 200\,$\mu$m at the dust temperature T$_{\rm d}$. 

Altogether,  we derived 26 sets of T$_{\rm d}$, I$_{200}$/A$_{\rm V}$ and 
$\tau_{200}$/A$_{\rm V}$ values in the 22 maps, 
due to multiple temperature components in 4 fields.
In addition, six fields of the del Burgo et al. (2003) sample have been 
reprocessed following our scheme (see Table~1).  

The errors of \tauav{200} come from two sources: from the error in the
determination of \iav{200} in the I$_{200}$ versus $A_{\rm V}$ scatter plot
and from the temperature uncertainty in B$_{\nu}\rm(T)$. Assuming that these
two sources are independent, the final $\tau_{200}$/A$_{\rm V}$ errors can be
expressed via the partial derivatives by \iav{200} and B$_{\nu}\rm(T)$,
following standard error propagation.
Since we use the slope of the I$_{200}$ versus A$_{\rm V}$ relation only,  systematic
offsets in I$_{200}$ and A$_{\rm V}$  -- like in the temperature computation --  do
not play a role. 

To determine the propagation of the error in T$_{\rm d}$ in the Planck
function, we use the two values T--$\delta$T and T+$\delta$T to estimate
the upper and the lower error bars in $\tau_{200}$/A$_{\rm V}$. In most cases
the upper and lower error bars are nearly similar, therefore we give one
average value for the uncertainty in Table~1. In the
subsequent figures we present these upper and lower error bars
individually. The relative errors caused by the temperature uncertainties
are typically much larger than the relative errors from the determination
of \iav{200} in the scatter plots.  

%%%%%%%%%%%%%%%%%%%%%%%%%%%%%%%%%%%%%%%%%%%%%%%%%%%%%%%%%%%%%%%%%%%%%%%%%%%%%%%%%%%%%%%%%%%%%%%%%%%%%%%%%%%
%%%%% RESULTS
%%%%%%%%%%%%%%%%%%%%%%%%%%%%%%%%%%%%%%%%%%%%%%%%%%%%%%%%%%%%%%%%%%%%%%%%%%%%%%%%%%%%%%%%%%%%%%%%%%%%%%%%%%%
\subsection{Emissivity versus temperature relationships}

%%%%%%%%%%%%%%%%%%%%%%%%%%%%%%%%%%%%%%%%%%%%%%%%%%%%%%%%%%%
\begin{table*}
\begin{tabular}{lrrrrrrr}
\hline
field & ISO-id & $\alpha_{J2000}$ &  $\delta_{J2000}$ & $\lambda_{short}$/$\lambda_{long}$ & T$_d$ & I$_{200}$/A$_V$ & $\tau_{200}$/A$_V$  \\
  &  [$\lambda_{short}$ / $\lambda_{long}$] &  (h~m~s) & (\degr ~\arcmin ~\arcsec) & ($\mu$m) & (K) & (MJy\,sr$^{-1}$\,mag$^{-1}$)  & ($\times$10$^{-4}$\,mag$^{-1}$)  \\
\hline
G004.3+35.8 & 10101158/10101157 & 15 53 46 & -4 31 19 & 100 / 200 & 14.0$\pm$0.2 &  11.3$\pm$  0.8 &    3.9$\pm$  0.3 \\
G100.0+14.8 & 40100614/39602313 & 20 32 45 & 65 19 49 &  90 / 200 & 16.0$\pm$0.1 &  10.7$\pm$  0.5 &    1.9$\pm$  0.1 \\
G101.8+17.0 & 80000751/80000652 & 20 23 19 & 68 01 55 & 100 / 200 & 14.7$\pm$0.4 &   5.2$\pm$  1.6 &    1.4$\pm$  0.4 \\
G102.0+15.2 & 40100818/40100717 & 20 40 06 & 67 10 43 &  90 / 200 & 15.0$\pm$0.0 &  10.8$\pm$  0.4 &    2.6$\pm$  0.1 \\
G114.0+14.9 & 75501106/75400905 & 22 28 06 & 75 13 22 & 120 / 200 & 13.3$\pm$0.3 &   8.0$\pm$  1.4 &    3.6$\pm$  0.6 \\
G114.0+14.9 & 75501106/75400905 & 22 28 06 & 75 13 22 & 120 / 200 & 12.5$\pm$0.3 &   4.4$\pm$  0.9 &    2.8$\pm$  0.5 \\
G114.3+14.7 & 75501211/75501210 & 22 33 03 & 75 14 38 & 120 / 200 & 12.1$\pm$0.6 &   4.2$\pm$  0.5 &    3.3$\pm$  0.4 \\
G114.6+14.6 & 76701109/76701108 & 22 37 45 & 75 13 13 & 120 / 200 & 12.9$\pm$0.1 &   6.7$\pm$  0.5 &    3.5$\pm$  0.3 \\
G114.6+14.6 & 76701109/76701108 & 22 37 45 & 75 13 13 & 120 / 200 & 12.2$\pm$0.1 &  10.1$\pm$  0.9 &    7.4$\pm$  0.7 \\
G121.6+24.6 & 56201502/56201701 & 23 07 27 & 87 10 30 &  90 / 200 & 15.5$\pm$0.1 &   8.0$\pm$  0.5 &    1.7$\pm$  0.1 \\
%G122.0+24.2 & 36803005 & 23 43 45 & 86 58 18 &  90 / 200 & 14.9$\pm$0.1 & 
%   3.5$\pm$  0.6 &    0.9$\pm$  0.2 \\
G122.0+24.2 & 56201606/36803005 & 23 43 45 & 86 58 18 &  90 / 200 & 14.9$\pm$0.1 &    3.5$\pm$  0.6 &    0.9$\pm$  0.2 \\
G142.0+38.5 & 15600853/15600851 &  9 32 57 & 70 26 12 & 100 / 200 & 15.1$\pm$0.2 &    3.4$\pm$  0.6 &    0.8$\pm$  0.2 \\
G170.2$-$16.0 & 85701512/85701411 & 4 21 20 & 27 00 42 & 120 / 200 & 13.3$\pm$0.3 &    6.3$\pm$  1.1 &    2.8$\pm$  0.5 \\
G170.2$-$16.0 & 85701512/85701411 & 4 21 20 & 27 00 42 & 120 / 200 & 12.4$\pm$0.3 &    3.8$\pm$  1.0 &    2.5$\pm$  0.7 \\
G173.9$-$15.7 & 68200305/68200403 & 4 32 41 & 24 36 18 & 120 / 200 & 12.9$\pm$0.2 &    5.7$\pm$  0.5 &    3.0$\pm$  0.3 \\
G174.3$-$15.9 & 68400606/68301104 & 4 32 58 & 24 10 51 & 120 / 200 & 13.2$\pm$0.1 &    6.0$\pm$  0.3 &    2.8$\pm$  0.1 \\
%G295.3$-$36.1 & 71601409 & 3 06 42 & -78 54 08 & 120 / 200 & 14.6$\pm$0.2 &   3.0$\pm$  1.8 &  0.8$\pm$  0.5 \\
G297.3$-$16.2 & 26100704/26100703 & 11 04 19 & -77 51 12 & 100 / 200 & 14.4$\pm$0.1 &  5.6$\pm$  0.3 &    1.7$\pm$  0.1 \\
G300.2$-$16.8 & 15701656/15701655 & 11 53 21 & -79 21 41 & 120 / 200 & 14.2$\pm$0.1 & 18.4$\pm$  0.5 &    5.9$\pm$  0.2 \\
G301.7$-$16.6 & 14102209/14102110 & 12 25 24 & -79 22 47 &  90 / 200 & 15.2$\pm$0.2 & 14.7$\pm$  1.1 &    3.4$\pm$  0.2 \\
%G302.6$-$15.9 & 27600419 & 12 44 55 & -78 48 24 & 100 / 200 & 17.8$\pm$0.7 & 2.5$\pm$  1.0 &  0.3$\pm$  0.1 \\
G302.6$-$15.9 & 27600420/27600419 & 12 44 55 & -78 48 24 & 100 / 200 & 13.8$\pm$0.3 & 9.8$\pm$  0.9 &    3.6$\pm$  0.3 \\
G303.5$-$14.2 & 71801760/33300559 & 13 00 47 & -77 06 09 & 100 / 200 & 16.8$\pm$0.1 & 6.5$\pm$  1.2 &    0.9$\pm$  0.2 \\
G303.5$-$14.2 & 71801760/33300559 & 13 00 47 & -77 06 09 & 100 / 200 & 13.3$\pm$0.1 & 9.8$\pm$  0.5 &    4.5$\pm$  0.2 \\
%G303.8$-$14.2 & 71901119 & 13 07 04 & -77 01 02 & 100 / 200 & 15.8$\pm$0.1 & 3.8$\pm$  1.1 &    0.7$\pm$  0.2 \\
G303.8$-$14.2 & 71901220/71901119 & 13 07 04 & -77 01 02 & 100 / 200 & 14.7$\pm$0.1 & 5.1$\pm$  0.9 &    1.4$\pm$  0.2 \\
%G353.3+16.7 & 26701013 & 16 27 48 & -24 25 39 & 100 / 200 & 16.5$\pm$0.5 & 46.5$\pm$  2.9 &    7.3$\pm$  0.5 \\
G355.3+14.7 & 31000632/31000631 & 16 40 03 & -24 11 48 & 100 / 200 & 15.6$\pm$0.5 & 5.7$\pm$  1.1 &    1.1$\pm$  0.2 \\
G359.1+36.7 & 43100630/43100629 & 15 40 09 & -7 12 29 & 100 / 200 & 16.1$\pm$0.1 & 13.0$\pm$  0.5 &    2.3$\pm$  0.1 \\
%G359.8$-$18.4 & 33401343 & 19 04 02 & -37 15 30 & 100 / 200 & 14.4$\pm$0.3 & 2.4$\pm$  0.7 &    0.7$\pm$  0.2 \\
G359.9$-$17.9 & 33401134/33401133 & 19 02 18 & -36 59 31 & 100 / 200 & 15.4$\pm$0.4 & 8.6$\pm$  1.6 &    1.8$\pm$  0.3 \\
\hline
G089.0$-$41.2$^*$ & 21600513/21600514 & 23 08 35 & 14 45 45 &  90 / 200 & 16.3$\pm$0.2 & 12.8$\pm$  3.9 &    2.1$\pm$  0.6 \\
%G090.7+38.0$^*$ & 27701701 & 16 50 53 & 60 55 26 & 100 / 200 & 18.7$\pm$0.2 & 3.3$\pm$  0.4 &    0.3$\pm$  0.0 \\
G111.2+19.6$^*$ & 11100606/11101606 & 21 02 16 & 76 51 50 & 150 / 200 & 15.2$\pm$0.2 & 13.9$\pm$  1.5 &    3.1$\pm$  0.3 \\
G187.3$-$16.7$^*$ & 82901031/82901031 & 5 02 24 & 13 41 15 & 120 / 200 & 14.6$\pm$0.1 & 12.2$\pm$  0.8 &    3.3$\pm$  0.2 \\
G297.3$-$15.7$^*$ & 26101501/26101401 & 11 07 55 & -77 28 15 & 150 / 200 & 14.7$\pm$0.1 & 7.0$\pm$  0.4 &    1.9$\pm$  0.1 \\
G301.2$-$16.5$^*$ & 60601027/60601027 & 12 16 22 & -79 17 09 & 120 / 200 & 13.2$\pm$0.1 & 8.9$\pm$  1.0 &    4.1$\pm$  0.4 \\
G301.7$-$16.6$^*$ & 60600925/60600925 & 12 25 24 & -79 22 47 & 120 / 200 & 12.8$\pm$0.1 & 19.4$\pm$  3.1 &   10.5$\pm$  1.7 \\
\hline
\end{tabular}

\label{table:summary}
\caption{Summary table of the basic properties and derived quantities
of the fields analysed in this study. The columns of the table are:
(1) the name of the field (denoting central galactic coordinates);
(2) ISO-id (TDT-number, Laureijs et al., 2003) of the short and 
long wavelength measurements. 
(3--4) right ascension and declination of the field centres (J2000);
(5) short and long ISOPHOT filter wavelengths used;
(6) dust temperature derived from ISOPHOT short- versus long wavelenght 
(200\,$\mu$m) surface brightness scatter plots, the uncertainties
presented here reflect the formal errors of the fits only;
(7) fitted slope of the 200\,$\mu$m surface brightness (I$_{200}$) versus
visual extinction (A$_{\rm V}$) scatter plot;
(8) dust emissivity derived from columns \#6 and \#7.}
\end{table*}
%%%%%%%%%%%%%%%%%%%%%%%%%%%%%%%%%%%%%%%%%%%%%%%%%%%%%%%%%%%

\begin{figure}
\includegraphics[width=8.5cm]{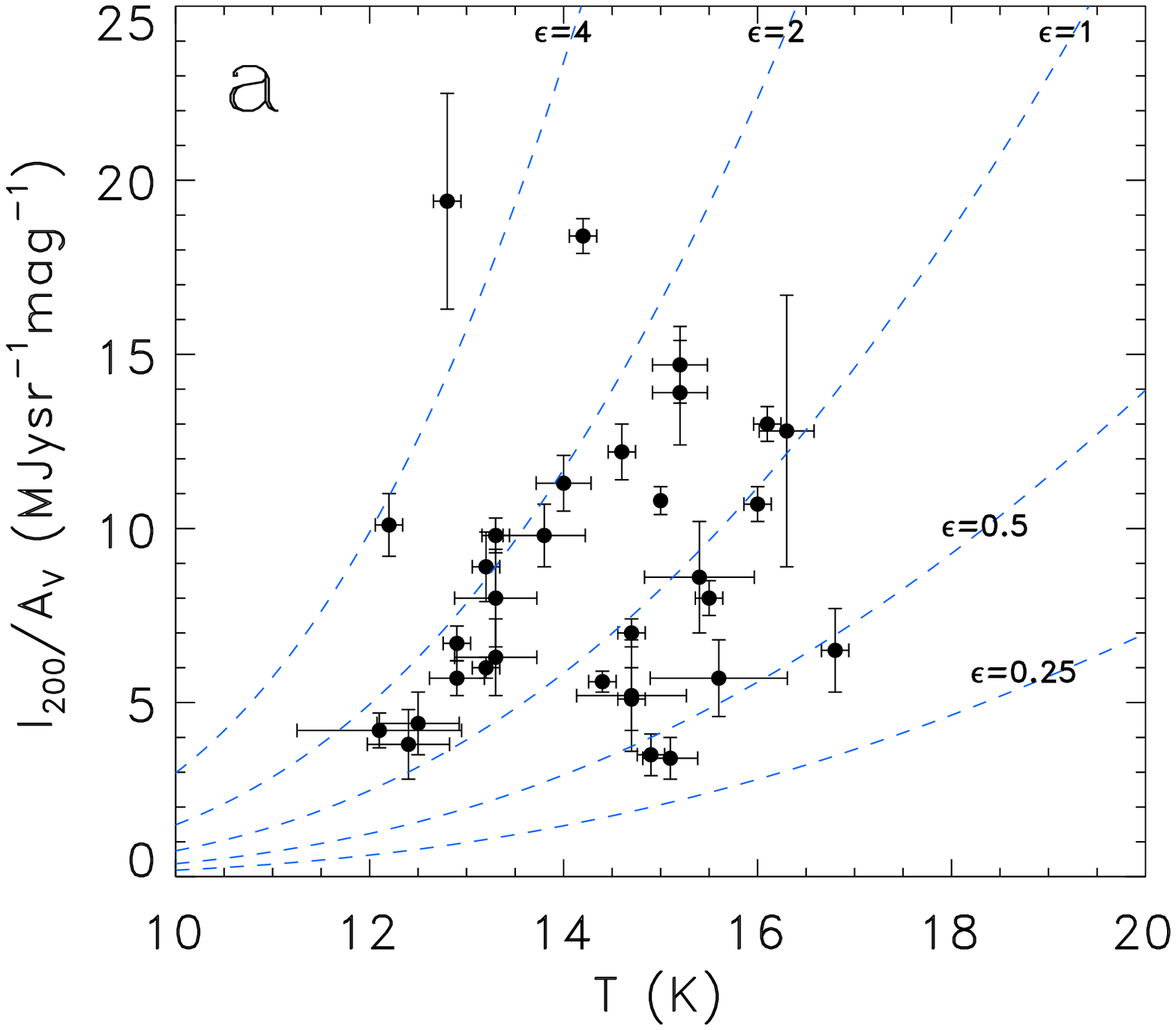}\\
\includegraphics[width=8.5cm]{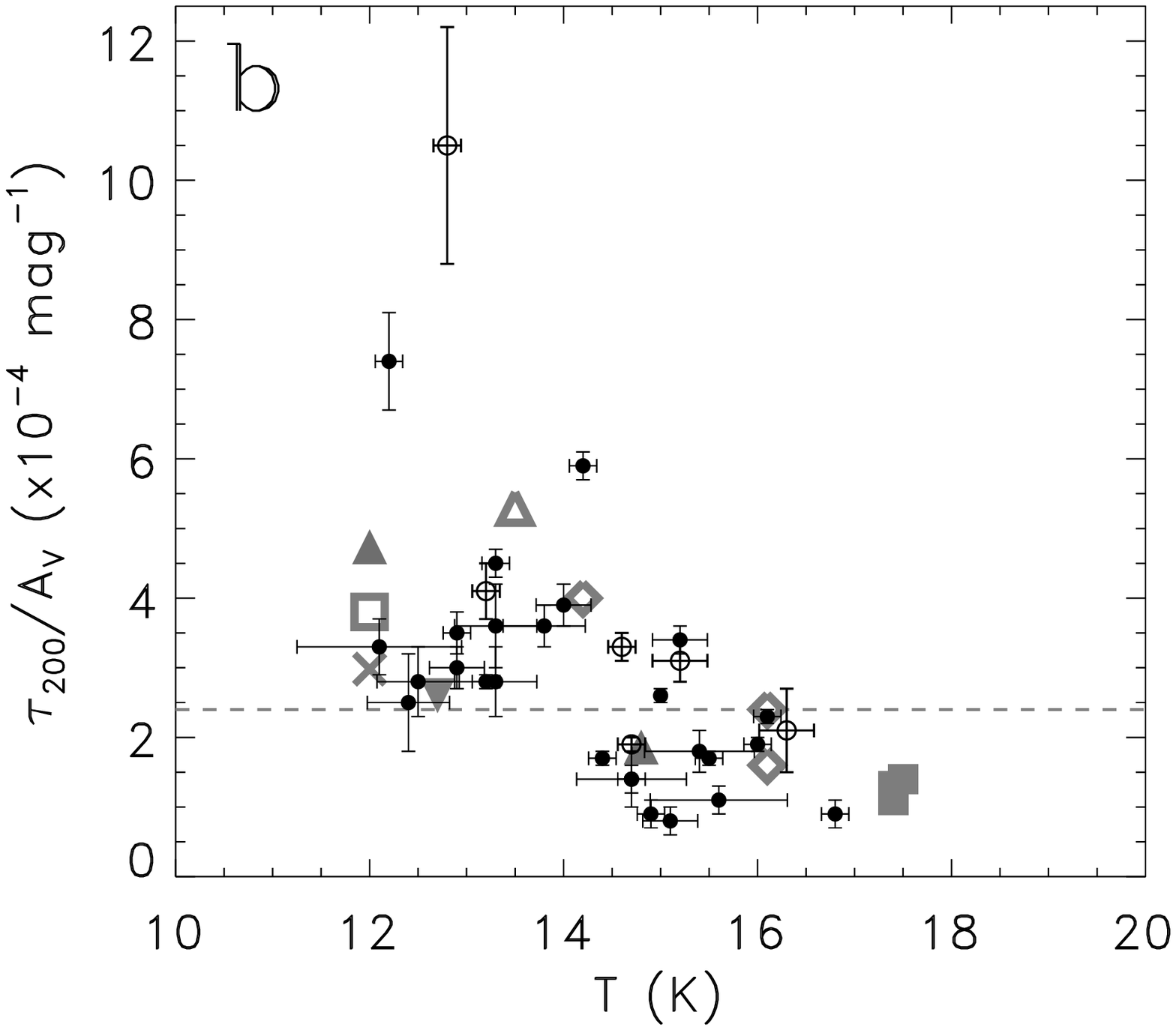}
\caption[]{
({\bf a}) Relationship between I$_{200}$/A$_{\rm V}$ and the dust temperature. 
Dashed lines represent iso-emissivity 
($\epsilon$\,=[{\tauav{200}}]/[{\tauav{200}]$_{\rm DISM}$}\,=\,constant) curves. 
The corresponding emissivity values are indicated.
({\bf b}) Relationship between the 200\,$\mu$m emissivity 
($\tau_{200}$/A$_{\rm V}$) and the dust colour temperature. Black dots with 
error bars mark the fields investigated
in this study. Open circles with error bars mark the 
reprocessed del Burgo et al. (2003) target fields. 
We also present recent results published in other papers, marked
by gray symbols: 
-- filled triangles: Stepnik et al. (2003); 
-- filled, upside-down triangle: Pagani et al. (2003);
-- filled squares: Rawlings et al. (2005);
-- cross: Bianchi et al. (2003);
-- open triangle: Lehtinen et al. (1998);
-- open diamonds: Lehtinen et al. (2004);
-- open square: Juvela et al. (2002). }
\label{fig:res}
\end{figure}

In Fig.~\ref{fig:res} we present the resulting ratios of I$_{200}$/A$_{\rm V}$
and $\tau_{200}$/A$_{\rm V}$ as a function of T$_{\rm d}$ for our sample of 
targets, including the reprocessed del Burgo et al. (2003) sample.
%In order to compare our values to that of the
%diffuse interstellar matter we defined an \emph{emissivity ratio} parameter as 
%$\epsilon$\,=[{\tauav{200}}]/[{\tauav{200}]$_{\rm DISM}$} 
%(see Fig.~2a). 
In addition, we included in Fig.~\ref{fig:res}b the values obtained by
different authors (for references see figure caption). The I$_{200}$/A$_{\rm V}$
ratios of the different studies were made consistent by converting to
$\lambda$= 200\,$\mu$m assuming a modified black-body with a
$\beta$\,=\,2 emissivity law. 
The \tauav{200} values from other authors, 
which were originally obtained at longer wavelengths, were transformed 
to 200\,$\mu$m as well. A comparison of our results with these previous 
works is presented in Sect.~5.1.

In Fig.~\ref{fig:res} the dashed line marks the value of the dust
emissivity  representative of the DISM. Note, that the DISM itself has a
unique temperature of $\sim$17.5\,K. 
The observed values of $\tau_{200}$/A$_{\rm V}$ exhibit significant deviations
from this  reference value as a function of temperature. Fields above a
dust temperature of 14\,K show lower emissivity value than that of the
DISM with no obvious trend in their distribution. In the colder fields
emissivity values are above that of the DISM, with a weak indication of a
decreasing trend towards lower temperatures. 

A few points are above the typical emissivity values
found for the majority of the fields at low temperature. 
The higher uncertainties in these fields -- as derived from the
uncertainites of the temperature and \iav{200} ratio determination --
cannot explain the high \tauav{200} values observed. 
Further potential sources of systematic errors that
could lead to incorrect emissivity values are discussed in detail
in Sect.~4.

%%%%%%%%%%%
%%%%%%%%%%
%%%%%%%%%%%%%%%%%%%%%%%%%%%%%%%%%%%%%%%%%%%%%%%%%%%%%%%%%%%%%%%%%%%%%%%%%%%%%%%%%%%%%%%%%%%%%%

\section{Uncertainties in the computation of emissivity parameters}

The reliability of the final emissivity values derived for our target
fields may be effected by several sources of errors. 
This is specially true, 
if a certain source of error does not contribute only to the 
random scatter of the derived values, but causes a certain pattern,
which may be mistaken for a physical change in dust properties. 
In this section we investigate several potential
error sources, and try to estimate their impact on the final 
emissivity determination. 

\subsection{Derivation of the correct extinction value
\label{sect:error_extinction}}

The extinction value A$_{\rm V}$ depends on the wavelength regime and the
method used to derive it. In Sect.~2.2 we compared the results from the
USNO star count and the  near-infrared NICER method and concluded that the
latter one is more suitable for our study. 

The presence of foreground stars may also have an effect on the derived
visual extinction values. While in a low extinction part of the cloud
most stars in the beam are background objects, towards the densest peaks
the stellar sample in the beam may be dominated by foreground stars. As a
result, the visual extinction in the densest parts can be underestimated,
and the emissivity value overestimated.
Since the foreground objects show low E(B-V) values, we minimized the
bias they cause by using an outlier resistant method to calulate the
average reddening value in each cell in the NICER method. However, all of
our target clouds are nearby (closer than a few hundred parsecs) and are
located at high Galactic latitudes (b\,$\ge$\,15\degr), therefore the
contribution of foreground stars is negligible. 

The extinction value may also depend on the size of the grid and whether
it is of fixed or adaptive type grid. Calculation of extinction/reddening
using an adaptive gridsize may significantly underestimate the
extinction, especially in a cloud centre, due to a inhomogeneous
distribution of stars in the cells. We therefore preferred to use a
fixed gridsize in our analysis.

\subsection{Dependency of R$_{\rm V}$}

The ratio of total over selective extinction R$_{\rm V}$ is found to be
significantly higher in dense interstellar clouds than R$_{\rm V}$ in
the diffuse ISM (Cardelli et al., 1989; Whittet et al., 2001). This
change is a strong indicator that the optical properties of dust grains
differ in dense regions. When we derived A$_{\rm V}$ from A$_J$ by following
the method of Lombardi \& Alves (2001), we adopted 
R$_{\rm V}$\,=\,3.1, which is a
value typical for the diffuse ISM and may not be representative for the
regions we investigate.   

% In dense clouds, which are typically of lower dust temperature, R$_V$ is
% usually higher (\lele{4}{Rv}{5}) than our adopted value of 3.1.
Applying the empirical relationship between R$_{\rm V}$ and 
$\rm {A_J}/{A_V}$ as
determined by Cardelli et al (1989), we find $\rm {A_J}/{A_V}$\,=\,0.282 and
0.334, for R$_{\rm V}$\,=\,3.1 and 5.5, respectively, implying that $A_V$ will
decrease by 19\% when increasing R$_{\rm V}$ from values representative of the
diffuse ISM to dense lines of sight. The corresponding \iav{200} and
$\tau_{200}$/A$_{\rm V}$ values will increase accordingly by 19\%.
%In these clouds A$_{\rm V}$ decreases by 15--20\%, 
%and the corresponding I/A$_{\rm V}$ and $\tau$/A$_{\rm V}$ 
%values increase accordingly by 15--20\%. 
We infer that the uncertainty in the assumption of R$_{\rm V}$\,=\,3.1 in the
deter\-mination of $A_V$ has a relatively minor implication on the values
of the observables presented in this paper. The assumption may introduce an
under\-estimation of at most 20\% in both \iav{200} and $\tau$/A$_{\rm V}$.

\subsection{Application of a different emissivity law \label{sect:errors_beta}}
Assuming that the exponent of the emissivity law $\beta$ is close to 2 in
the interstellar medium (Gezari et al. 1973, Draine \& Lee, 1984,
Boulanger et al., 1996), we adopted $\beta$=2 to derive T$_{\rm d}$ from the
slope of the short (90, 100 or 120\,$\mu$m) versus long (200\,$\mu$m)
wavelength scatter plots (Sect.~3.1).
Based on balloon-borne submillimetre observations of a large number of ISM
regions, Dupac et al. (2003) found that $\beta$ is not constant but
depends on the dust temperature as
$\beta(\rm T)\,=\,(0.4 + 0.008\cdot(\rm T/1\,\rm K))^{-1}$.
We used this relationship to investigate the effect of possible changes
in $\beta$ on our results. We calculated the new colour temperature from
the I$_{\rm short}$ versus I${_{200}}$ scatter diagrams using $\beta(T)$.
From this temperature we obtained the modified values for 
$\tau_{200}$/A$_{\rm V}$.

%%%%%%
\begin{figure}
\includegraphics[width=8.5cm]{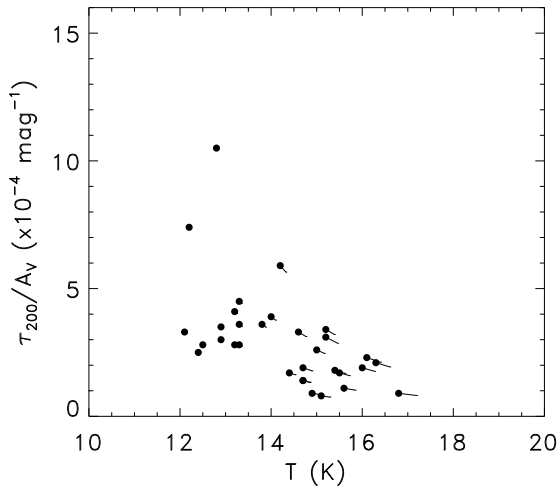}
\caption[]{Displacement of $\tau$/A$_{\rm V}$ vs. T$_{\rm d}$ data 
points due to the
effect of a temperature dependent $\beta$ (Dupac et al., 2003). The
original emissivity values with $\beta$\,=\,2 are marked by filled 
circles. The stripes originating from these points indicate the change
due to a different $\beta$.}
\vskip 0.5cm
\label{fig:dupac}
\end{figure}
%%%%%%%%%%%
%
In Fig.~\ref{fig:dupac} we show the effect of $\beta(T)$
on the temperature versus emissivity relationship.  
The $\beta$ dependency tends to increase the temperature,
but the change is small, only for temperatures higher than  15.5 K the
variations are noticeable, more than 0.3 K but less than 0.5 K. 
Although the new temperature-dependence of $\beta$ cause a systematic change of the
distribution of the measurements in Fig. \ref{fig:dupac}, the effect is
small with respect to the measurement uncertainties.

%%%%%%%%%%%%%%%%%%%%%%%%%%%%%%%%%%%%%%%%%%%%%%%%%%%%%%%%%%%%%%%%%%%%%%%%%%%%%%%%
\subsection{Presence of warm dust}

IRAS related studies of the dense and diffuse ISM have shown that
extended emission at 60 $\mu$m can be associated with dust in low density
regions. In these regions the 100 $\mu$m emission is closely correlated
with the 60 $\mu$m emission, suggesting that they trace the same dust
component in the ISM.
At higher densities the 60 $\mu$m emission becomes weak indicating the
absence of grains giving rise to the 60 $\mu$m emission. The remaining
100 $\mu$m emission is a signature of a ``cold'' dust component
at $T<$15 K. This observation is supported by the close correlation of
the cold dust component with $^{13}$CO emission.
The surface brightness measured at a wavelength of
$\lambda$\,$\approx$\,100$\mu$m may contain emission from both the cold
and the ``warm'' (or low density) dust component. The warm
component will affect the determination of the $I_{\nu}/I_{200}$ colour
temperature, such that the temperatures will be overestimated due to the
short wavelength emission component.

%This contribution can determined by the separation of the warm and cold
%components, using e.g. IRAS/ISSA (...) 60 and 100$ \mu$m surface brightness
%images (Laureijs et al....). The idea is that the warm component of the
%100$\mu$m  emission is correlated with the 60$\mu$m emission while the cold
%component does not.

\begin{figure}
\includegraphics[width=8.5cm]{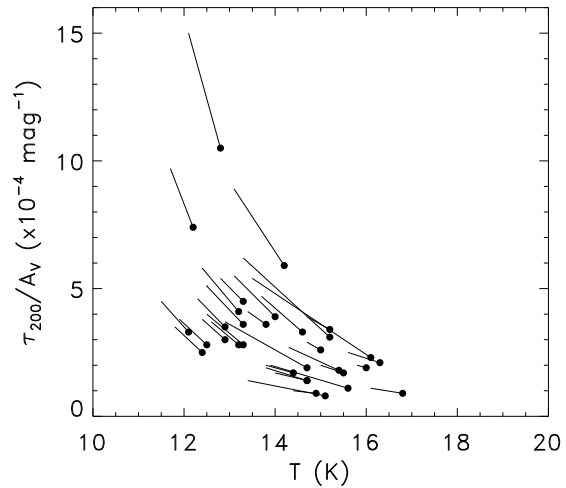}
\caption[]{Displacement of the T$_{\rm d}$--\tauav{200} data points 
with the correction for the warm dust component of 
the short wavelength surface brightness values. The original
data points are marked by filled circles and the 
stripes originating from them indicate the change due to 
the correction for the presence of warm dust. }
\vskip 0.5cm
\label{fig:res_icold}
\end{figure}

To determine the fraction of the cold component in our short wavelength 
($\sim$100\,$\mu$m) maps we analysed 60 and 100\,$\mu$m maps of our regions
obtained with IRAS/ISSA because in most fields there are no ISOPHOT
observations at 60\,$\mu$m. 
%60 and 100\,$\mu$m surface brightness values were transferred to
%the COBE/DIRBE photometric
%system using the conversion factors in Wheelock et al. (1994). 
The amount of emission from the cold component at 100 $\mu$m towards a
specific sky position can be calculated from the two IRAS bands:
\begin{equation}
I_{cold} = I_{100} - \theta \cdot I_{60} 
\end{equation}
where $\theta$ is the slope of a linear relationship fitted to the 60 vs.
100$\mu$m surface brightness correlation diagram in the outer parts of
the cloud. 
The relative contribution of the cold dust component at 100 $\mu$m is
estimated from the correlation between I$_{cold}$ and the original
I$_{100}$ surface brightness. The scatter diagram was fitted by a linear
function, resulting in a slope $X_{100}$, which gives the ratio of the
cold dust emission to total surface brightness at 100 $\mu$m. We find
typical values of \lele{0.7}{$X_{100}$}{0.8}.  
%Due to the possible calibrational discrepancies between the ISOPHOT and
%IRAS/ISSA photometric systems, 
We used $X_{100}$ to determine the cold dust emission contribution in the
ISOPHOT short wavelength bands.
%rather than subtracting $\theta \cdot I_{60}$ directly from the ISOPHOT maps. 
%
Since the ISOPHOT measurements were not always taken at 100 $\mu$m, we
converted $X_{100}$ to $X_\lambda$ assuming that the spectral energy
distribution of the total surface brightness can be approximated by two
modified black-bodies ($\beta$\,=\,2) with two different temperatures,
T$_{warm}$ and T$_{cold}$. We fixed T$_{warm}$ to 17.5K (Lagache et al.,
1998) and set T$_{cold}$ to the temperature obtained from the original
surface brightness scatter plots. Our analysis shows that
$X_\lambda$/$X_{100}$ is close to 1 and the ratio is not very sensitive
to T$_{cold}$.
These $X_{\lambda}$ values were applied to the short wavelength surface
brightness values at 90, 100 or  120\,$\mu$m, to compute the corrected
$\tau_{200}$/A$_{\rm V}$ values. The results are presented 
in Fig.~\ref{fig:res_icold}. 
%%%%%%
Fig. \ref{fig:res_icold} shows that the presence of warm dust tends to
overestimate the temperature by 0.5--1\,K which is significant. As a
result, due to the lower temperature of the cold dust and since
$I_{200}$/A$_{\rm V}$ is hardly affected, 
correction for warm dust emission increases 
warm dust increases $\tau_{200}$/A$_{\rm V}$
by at most $30$\%.

%%%%%%%%%%%%%%%%%%%%%%%%%%%%%%%%%%%%%%%%%%%%%%%%%%%%%%%%%%%%%%%%%%%%%%%%%%%%%%%%%%%%%%%%%%%%
%%
\subsection{Effect of surface brightness calibration errors}
%%

%%%%%%
\begin{figure}
\includegraphics[width=8.5cm]{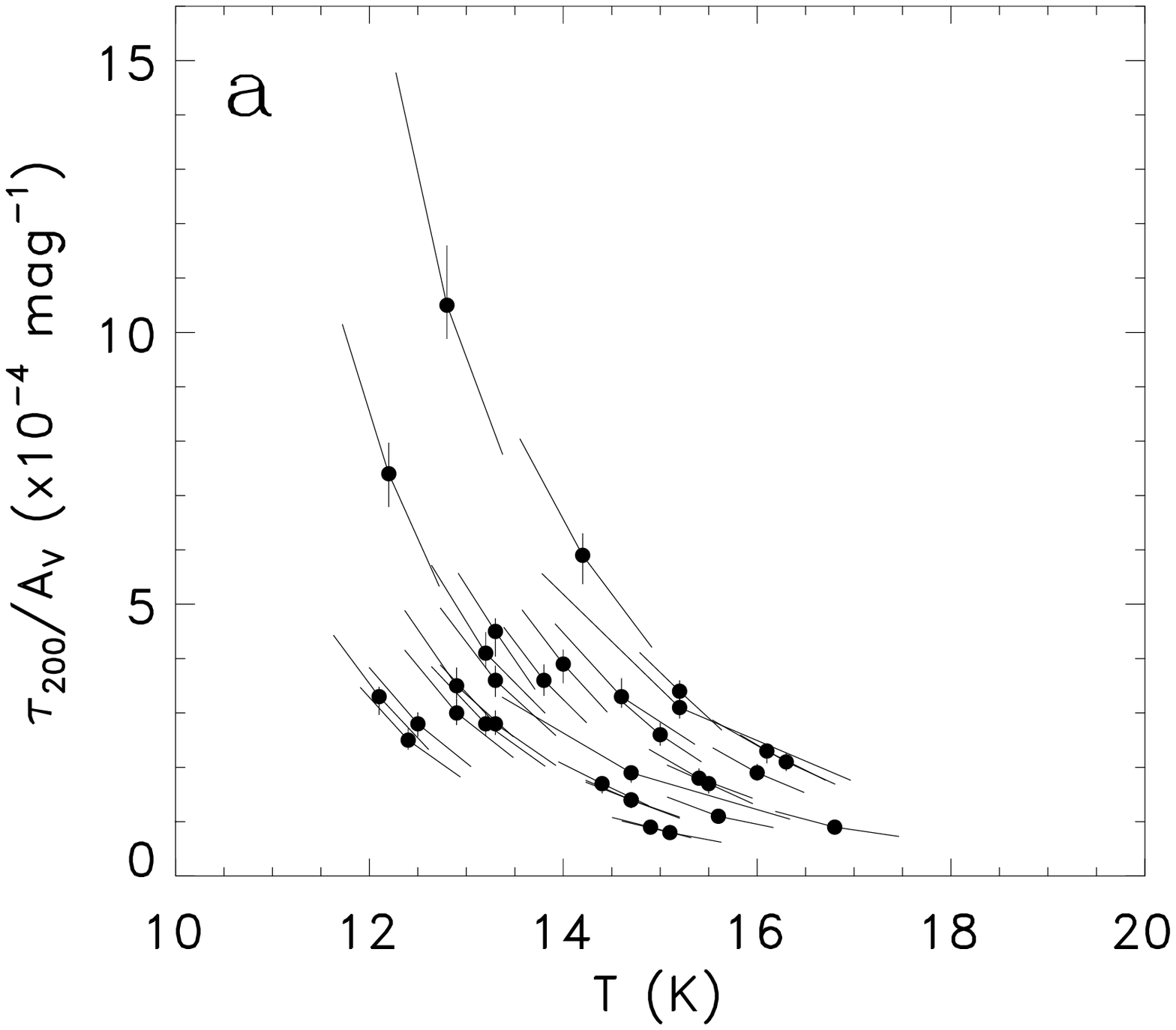}
\includegraphics[width=8.5cm]{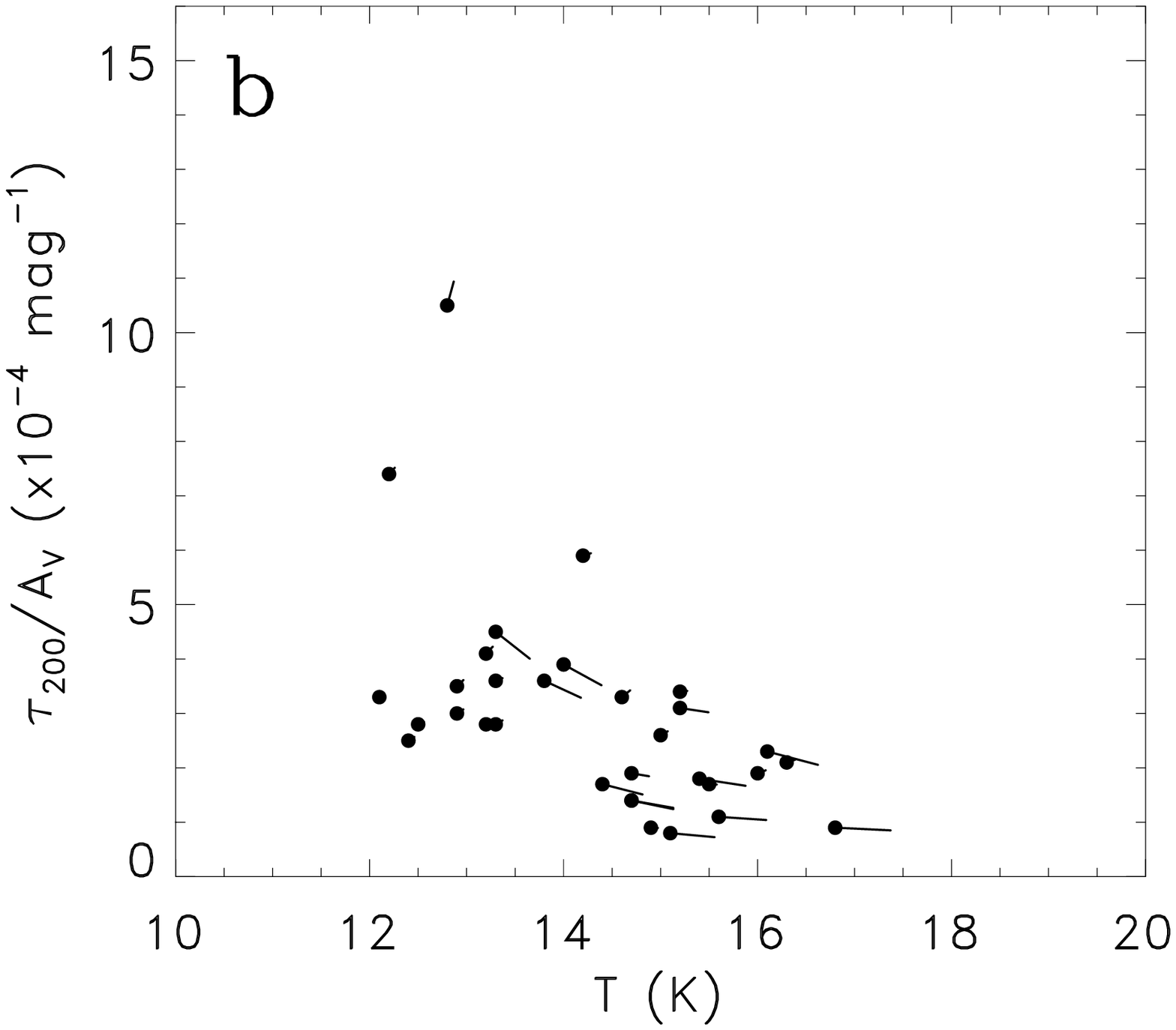}
\caption[]{Effect of calibration errors in the surface 
brightness calibration. 
({\bf a}) Statistical errors in the short and long wavelength surface brightness 
(see the text for details).
The four lines originating from each black dot (the
original \tauav{200} data points) span 
a banana-shape area in the figure, corresponding to 9\% calibration
errors in both the short and long wavelength ISOPHOT surface brightness
values, as found in Appendix~A. 
({\bf b}) Displacement of the \tauav{200} values (stripes) relative to 
the original ones (black dots, as listed in Table~1), 
when the correlation coefficient between the ISOPHOT and DIRBE 
surface brightness calibration (see Appendix~A) are applied.}
\vskip 0.5cm
\label{fig:systematic}
\end{figure}
%%%%%%%%%%%
%%%%%%%%%%%
%%

Errors in the ISOPHOT surface brightness cali\-bration may
affect the derived dust temperatures and the relationships presented
in Fig.~\ref{fig:res}. Here we first discuss the general 
effect of surface brightness calibration errors applying a simple model
in Sect.~4.5.1.

Actual calibrational errors in this photometric 
system can only be unraveled by a comparison with a standard
system. Due to its measurement design the DIRBE instrument on-board 
the COBE satellite was able to perform accurate absolute 
photometric observations (Hauser et al. 1998a) and serves now as the standard 
system for infrared sky brightness observations.  
A detailed comparison of the surface brightness photometric systems 
of ISOPHOT and that of COBE/DIRBE is presented in Appendix~A. 
In Sect.~4.5.2 we apply the ISOPHOT--DIRBE
surface brightness transformation equations found in Appendix~A, and discuss
their effect on our results.  

\subsubsection{General effect of calibration errors}

In order to test the effect of the surface brightness calibration
errors on the derived emissivities, in a simple approach we assumed
a linear relationship between the
true sky brightness and the surface brightness measured by a specific 
instrument/filter, i.e. the transformation equation between the two can 
be written as:
\begin{equation}
 \mathrm I_{\lambda}^{\rm meas}\,=\,{\rm G_{\lambda}}{\times}I_{\lambda}^{\rm sky} +
    {\rm C_{\lambda}} 
\end{equation}

where I$_{\lambda}^{\rm meas}$ is the 
measured sky brightness, I$_{\lambda}^{\rm true}$ is the
real sky brightness, and G (gain) and C (offset) are the coefficients 
describing  the transformation.  

Our analysis method,
which is based on slope fitting in scatter plots (see Sect.~2), is not
sensitive to errors due to constant offsets in the surface brightness. 
Multiplicative errors, 
however, change both the colour temperature and the
emissivity parameters. Such errors may come from the presence of a
gain uncertainty due to incorrect surface brightness calibration. 
We investigated the impact of this effect on our emissivity values, 
using a simple model. We assumed that
I$_{\rm short}^{\rm meas}$\,=\,G$_{\rm short}\cdot$I$_{\rm short}^{\rm sky}$ 
and I$_{200}^{\rm meas}$\,=\,G$_{200}\cdot$I$_{200}^{\rm sky}$, 
where I$_{\rm short}^{\rm sky}$ and
I$_{200}^{\rm sky}$ are the true surface brightness values of the sky
at our short wavelength (either 90, 100 or 120\,$\mu$m)
 and at 200\,$\mu$m, respectively, while
I$_{\rm short}^{\rm meas}$ and I$_{200}^{\rm meas}$ are the values 
affected by calibration errors. G$_{\rm short}$ and G$_{200}$ 
are the detector gains for the short wavelength and for the 200\,$\mu$m
filters, respectively. For both G$_{\rm short}$ and G$_{200}$ 
we assumed two extrema, 0.91 and 1.09, correspondig to an uncertainty
value of 9\%, as found for the ISOPHOT filters in Appendix~A. 

The effect of the four gain combinations for the derived emissivity values
and dust temperatures is presented in Fig.~5a. 
The shorter, fully vertical lines originated
from the black dots correspond to the cases, where both G$_{\rm short}$
and G$_{200}$ have the same value (either the same 0.91 or 1.09). 
In these cases the changes in the \tauav{200} values are 9\%.
   
The longer, tilted lines correspond to the cases, when G$_{\rm short}$
and G$_{200}$ have opposite values. Then the amount of change in 
\tauav{200} and T$_{\rm d}$ 
depends noticeably on the original temperature, but in general
is 25...30\%. This is a major effect, but the \tauav{200} values
remain in the domain, that their relative deviations from the
emissivity of the diffuse interstellar matter is still significant. 

The difference between the original and recalcultated dust 
temperatures are in the order of $\sim$0.7\,K. As discussed in 
Sect.~3.1 these calibration errors are the main sources of uncertainty 
in the temperature determination. 

The scenario presented here is indeed a 'worst-case', i.e. 
the relative effect of the calibration errors at the two 
wavelengths are the strongest. The likely 
effect is smaller and would result in a final \tauav{200}
uncertainty of 10...20\%, depending on the original colour temperature. 
Here we also did not take into account the systematic calibrational
differences between the two photometric systems, which could 
significantly reduce the final \tauav{200} uncertainties 
(see Sect.~4.5.2).  

%%%%%%%%%%%%%%%%%%%%%%%%%%%%%%%%%%%%%%%5

\subsubsection{Effect of transformation between the ISOPHOT and 
COBE/DIRBE surface brightness photometric systems}

We tested the effect of the correction of ISOPHOT surface brightness values
by the ISOPHOT--DIRBE transformation coefficients found in Appendix~A 
(see Table~\ref{table-corr}). Both the dust temperatures and the \iav{200} 
ratios have been recalculated by the corrected surface brightness values in
order to obtain the corrected \tauav{200} emissivities. The results are shown
in Fig.~5b. The effect of this correction is small for the \tauav{200} values
obtained from measurements with the 90 and 120\,$\mu$m ISOPHOT filters. 
The 100\,$\mu$m-related data points are more displaced, due to the 
relatively large difference in the scaling factors. However, the overall
shape of the distribution did not change significantly, and 
the effect is much smaller than that of the general calibration errors
presented in Sect.~4.5.1.

\subsection{The impact of various error sources on the emissivity -- 
dust temperature relationship}
The distribution of data points in Fig.~\ref{fig:res}b indicates a clear
trend in the \tauav{200} versus $T$ relationship: \tauav{200} is enhanced
for T\,$<$\,14\,K and decreases towards higher temperatures in the range
\lele{14}{T}{16}\,K. In the preceeding subsections we investigated
possible mechanisms which could explain this observation, our findings
are summarized below:

\begin{itemize}
  \item 2MASS data provide reliable extinction maps for
	our regions. Contamination by foreground stars -- which is most
	critical towards the densest regions --  cannot be excluded, but probably 
	has a minor effect. 
  \item The dense clouds in our sample, presumably also the cold ones, may have
	a ratio of total over selective extinction R$_V$ higher than the
	standard value of 3.1. As a consequence A$_{\rm V}$ may decrease 
	at most by $\sim$19\%, and the corresponding \iav{200} and 
	\tauav{200} values increase accordingly by the same fraction. 
	However, high R$_V$ should be restriced to small areas in our 
	target fields. These regions are probably excluded anyhow due to 
	their too low star count or too high extinction values 
	(A$_{\rm V}$\,$\ge$\,8\,mag). Therefore the likely effect of R$_V$ 
	on the final emissivity values is below $\sim$10\%.
  \item	Assuming a temperature dependent emissivity law $\beta(T)$ instead of a
	fixed $\beta$\,=\,2 has a minor effect on the \tauav{200}
	distribution (Fig.~\ref{fig:dupac}). The differences with
	$\beta$\,=\,2 are more prominent at higher temperatures, but the main
	trend for clouds colder than the dust in the DISM is not
	changed. 
  \item The possible presence of warm dust causes an overestimate of the dust
	temperature and, consequently, an underestimate of \tauav{200} by
	10--30\%. This variation is within the typical measurement
	uncertainties of the data points and does not alter the trend as
	observed in Fig.~\ref{fig:res}. 	
  \item Calibration errors can significantly modify the \tauav{200} 
         and dust temperature values. 
	The application of ISOPHOT surface brightness values 
	corrected for the systematic differences
	of the ISOPHOT and COBE/DIRBE surface brightness photometric
	systems does not change the observed distribution of \tauav{200}
	values significantly. Applying a $\pm$9\% general calibration
	uncertainty would cause a much severe effect,
	but the corrected results remain in the $\sim$25\% 
	range to the original values. Transformation of the ISOPHOT 
	surface brightness values to the DIRBE photometric system 
	has only a minor effect on the T$_{\rm d}$ vs. 
	\tauav{200} relationship. 

\end{itemize}

From the uncertainties due to different error sources we estimated 
a general $\sim$30\% uncertainty for our \tauav{200} values. 
%The uncertainties discussed
%above have the most severe impact on the high 200\,$\mu$m 
%emissivity values with low dust temperatures. 
To match the observations by assuming a constant emissivity
(that of the DISM) a 50...80\% uncertainty is required, which is 
unrealistic, according to our investigations. 
The observed temperature versus \tauav{200}
relationship, as presented in Fig.~\ref{fig:res}b, 
most probably indicates physical changes and is {\emph
not} due to artifacts in the data or to incorrect initial assumptions
in the calculations.

%%%%%%%%%%%%%%%%%%%%%%%%%%%%%%%%%%%%%%%%%%%%%%%%%%%%%%%%%%%%%%%%%%%%%%%%%%%%%%%
%% >>>>>>>>>>>>>>>>>>>>>> DISCUSSION <<<<<<<<<<<<<<<<<<<<<<<<<
%%%%%%%%%%%%%%%%%%%%%%%%%%%%%%%%%%%%%%%%%%%%%%%%%%%%%%%%%%%%%%%%%%%%%%%%%%%%%%%
\section{Discussion}

\begin{figure}
\includegraphics[width=8.5cm]{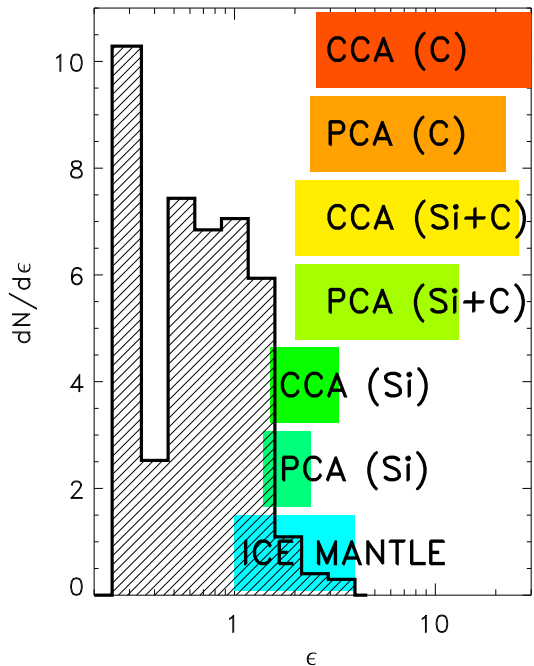}
\caption[]{The histogram with shaded area shows the relative
frequency ($dN/d\epsilon$) of the observed \tauav{200} values relative
to that of the DISM 
($\epsilon$\,=\,\tauav{200}/[\tauav{200}]$_{\rm DISM}$, where
[\tauav{200}]$_{\rm DISM}$\,=\,2.4$\cdot$10$^{-4}$\,mag$^{-1}$), 
found for our targets. Horizontal bars represent the predicted 
range of enhancement in emissivity for 
different types of grains following Stognienko et al. (1995), 
from bottom to top: (1) ice mantles, (2) silicate particle-cluster
aggregates, (3) silicate cluster-cluster aggregates, (4) mixed (silicate
and carbon) particle-cluster aggregates, (5) mixed cluster-cluster
aggregates, (6) carbon particle-cluster aggregates (7) carbon
cluster-cluster aggregates. }
\vskip 0.5cm
\label{fig:enhancement}
\end{figure}
%%%%%%%%%%%

\subsection{Observed variations in \tauav{200}}

%The regions published in the study by del Burgo et al. (2003) are
%included in the figure. To be consistent with our data set
%we have reprocessed these regions in the same way as the other fields in
%our study. 
We found notable differences between the original del Burgo et al. (2003) 
\tauav{200} values and the emissivities of the same fields reprocessed 
in this paper.  The main differences between our and the 
del Burgo et al. (2003) processing are that in this study 
we applied extinction values derived from 2MASS
data and that we used different short wavelength data for the temperature
calculation, namely 90, 100 or 120\,$\mu$m instead of 150\,$\mu$m (these
new results are also included in Table~1 and in Fig.~\ref{fig:res}). 
The dust temperatures we found are consistent with those in del Burgo et al. (2003).
However, the resulting \iav{200} ratios are different, and this discrepancy
lead to notably lower \tauav{200} values (see Table~1). This also indicates,
that it is mainly the application of 2MASS data that is responsible for the 
differences seen in the emissivities.

 Our values obtained for the TMC\,2 (G173.9--15.7 and G174.3--15.9)
and for LDN\,1780 (G359.1+36.7) are very similar to
the average values presented for these clouds by del Burgo \& Laureijs
(2005) and Ridderstad et al. (2006), respectively. In these papers
\iav{200} and \tauav{200} values were derived by taking the
brightness ratios from the maps instead of correlation plots.
Despite the general argeement with our data, the detailed analysis of
LDN\,1780 show that \tauav{200} can range from  10$^4$\,mag$^{-1}$ to
4$\cdot$10$^4$\,mag$^{-1}$ within the same cloud (Ridderstad et al.,
2006). An even higher increase of \tauav{200} is observed in the Taurus~2
molecular cloud (del Burgo \& Laureijs, 2005).

An important outcome of our analysis of a large sample of cloud regions
is that we find a clear trend of \tauav{200} versus $T_{\rm d}$, but that the
variation in \tauav{200} is less pronounced than initially reported by
other studies. Only three fields exhibit \tauav{200} which is more than 
threefold the value for the DISM (\tauav{200}$_{\rm DISM}$). 
All other regions with $T_{\rm d}<$14~K (12 regions) have 
\tauav{200}\,$>$\,\tauav{200}$_{\rm DISM}$. 
In the majority of the regions with $T_{\rm d}>$14\,K the 
emissivities are below \tauav{200}$_{\rm DISM}$ (10 out of 12 regions). 
One extreme case is G142.0+38.5, where 
3\tauav{200}\,$\approx$\,\tauav{200}$_{\rm DISM}$.

The values of \tauav{200} in our sample refer to the average properties
(T$_{\rm d}$, and \iav{200}) over a relatively extended region of 
$\sim$100\,arcmin$^2$. The enhancement observed by us of $\epsilon<2$ 
in most cases suggests, that higher $\epsilon$ values usually occur only in smaller 
regions inside the clouds, possibly in the densest cores.

We have not seen large deviations in the scatter plots, neither in the
\iav{200}, nor in the dust temperature derivation, as reflected
in the uncertanties in T$_{\rm d}$ and \iav{200}. Extremely high, local
emissivities are expected to show up in these scatter plots, and are
therefore excludeable in our fields at the resolution we adopted, namely
3\farcm5.  

\subsection{Changes of dust properties at low temperature}

Below 14\,K all \tauav{200} values are higher than
2.4$\cdot$10$^{-4}$\,mag$^{-1}$,  the representative value of the DISM. 
This finding agrees with the results of several earlier papers 
(see Sect.~1 and the caption of Fig.~2). 
The majority of our data points are within a
well-defined range, 1-to-2 times the emissivity of the DISM. Enhanced
emissivity values are usually interpreted as the change of dust
properties: coagulation of dust particles or formation of ice mantles on
grain surfaces (see e.g. Dwek 1997).
Dust grains may show a large variety both in composition (silicate,
carbon, or mixed) and in structure (ice mantle, particle-cluster
aggregates, cluster-cluster aggregates) and the different types show
different enhancement of the emissivity. Since the emissivity enhancement
typically is not higher than a  factor of 2, our results are consistent
with ice mantles or cluster of silicate particles (CCA or PCA, see
Fig.~\ref{fig:enhancement}). On the other hand far-infrared emissivities
produced by  grains containting carbon aggregates are too high for
most of our observed emissivity values. However, in some particular
regions the observed \tauav{200} exceeds significantly the
representative values of the DISM, and may indicate the existence of
carbonaceous grains.
{
%%%%%%%%
\begin{figure}
\includegraphics[width=8.5cm]{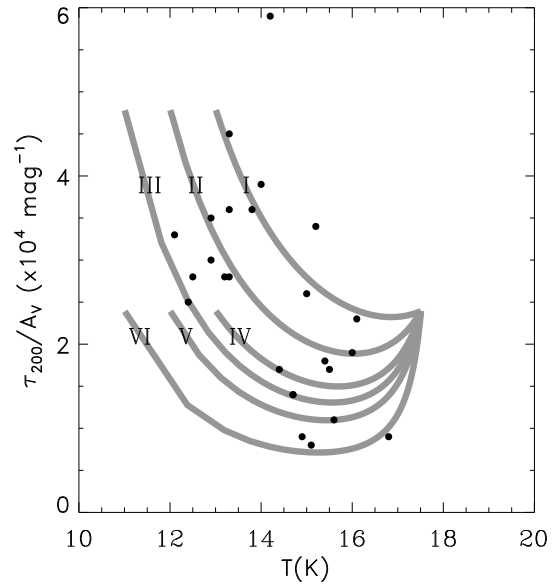}
\caption[]{In this figure we present model fits of T$_{\rm d}$ versus 
$\tau_{200}$/A$_{\rm V}$
with a dust emission of bimodal temperature distribution. 
The gray curves
represent different emissivities and cold temperatures (T$_{\rm c}$). 
To derive the "virtual" dust temperature, 
surface brightness at 100\,$\mu$m (short) and 
200\,$\mu$m (long) wavelengths were used. 
The warm dust temperature has been set to
T$_{\rm w}$\,=\,17.5\,K.  Other short wavelengths (90 or 120\,$\mu$) produce curves
very close to the 100\,$\mu$m ones.
{\bf I.} $\epsilon$\,=\,2.0, T$_{\rm c}$\,=\,13.0\,K;
{\bf II.} $\epsilon$\,=\,2.0, T$_{\rm c}$\,=\,12.0\,K;
{\bf III.} $\epsilon$\,=\,2.0, T$_{\rm c}$\,=\,11.0\,K;
{\bf IV.} $\epsilon$\,=\,1.0, T$_{\rm c}$\,=\,13.0\,K;
{\bf V.} $\epsilon$\,=\,1.0, T$_{\rm c}$\,=\,12.0\,K;
{\bf VI.} $\epsilon$\,=\,1.0, T$_{\rm c}$\,=\,11.0\,K; 
{\bf The black dots mark our measured \tauav{200} values.}}
\vskip 0.5cm
\label{fig:bimodal}
\end{figure}
%%%%%%%%%%%

\subsection{Emissivities lower than in the DISM}

As presented in Sect.~3.3 and Fig.~\ref{fig:res}, we find
low emissivity values for 14\,K\,$\le$\,T$_{\rm d}$\,$\le$\,16\,K colour
temperatures.  Similar results were obtained for some specific regions by
Lehtinen et al. (2004) and by Rawlings et al. (2005). 
Even though one can explain an emissivity lower than that of the DISM in terms
of changed grain properties, it is also possible
to explain this
behaviour using a bimodal temperature model (Cambresy et al., 2001; del
Burgo et al., 2003). In this model we assume, that the observed
far-infrared surface brightness is due to two emission components with
different dust temperatures, both due to big grains. The observed
infrared emission can be written as:
\begin{equation}
I_{\nu} = \tau_{\nu,c}B_{\nu}(T_{\rm c}) + \tau_{\nu,w}B_{\nu}(T_{\rm w})
\end{equation}
where T$_{\rm c}$ and $T_{\rm w}$ are the temperatures of the cold and warm dust
components,  respectively, and $\tau_{\nu,c}$ and $\tau_{\nu,w}$ are the
infrared optical  depths of the cold and warm components, respectively. 
The emissivity of the cold component may differ from that of the warm
component, and this change is characterized by the $\epsilon$ parameter:
\begin{equation}
\tau_{\nu,c} = \epsilon X \tau_{\nu,tot}
\end{equation} 
\begin{equation}
\tau_{\nu,w} = (1-X) \tau_{\nu,tot}
\end{equation} 
where $\tau_{\nu,tot}$ is the total effective optical depth if both
components had  identical emission properties and X is the fraction of
the optical depth of the  cold component with respect to the effective
total opacity  (for more details see del Burgo et al. 2003).  In del
Burgo et al. (2003) the cold temperature T$_{\rm c}$ was choosen to be
13.5\,K,  representing the lowest colour temperatures obtained from their
sample. However, in our sample we included denser molecular regions and
the presence of colour temperatures of  T$_{\rm d}$\,$\approx$\,12\,K obviously
requires a lower T$_{\rm c}$ value.  

We derived the colour temperature from the slope of the
scatter plot of two IR-wavelength surface brightness distributions. It is
easy to see that in a bimodal temperature distribution the variation of X
causes a change in the ratio of  surface brightness values measured at
two wavelengths (see Fig. 8 in del Burgo et al., 2003), i.e. an effect
very similar  to the change of the colour temperature. A specific X value
can be directly linked to a ``virtual'' colour temperature, as long as
$T_{\rm c}$ and $T_{\rm w}$ are fixed.
As we demonstrate below, mixing rather than
change of dust grain properties can be the reason that data points in
Fig.~\ref{fig:res}b fall below the representative value of the DISM.  
In Fig.~\ref{fig:bimodal} we plot $\tau_{200}$/A$_{\rm V}$ model curves with 
mixing ratios changing in the \lele{0}{X}{1} range, versus the
corresponding virtual colour temperature. In this model we used 
$\tau_{\nu,tot}$\,=\,2.4$\cdot$10$^{-4}$\,mag$^{-1}$, which is the
representative value of the DISM (Schegel et al., 1998). In all cases
we kept T$_{\rm w}$\,=\,17.5\,K and assumed different values of T$_{\rm c}$. For
$\epsilon$ either $\epsilon$\,=\,1 or $\epsilon$\,=\,2 was adopted.  

It is clear from Fig.~\ref{fig:bimodal}, that in order to reproduce the 
higher \tauav{200} values for cold regions (T\,$<$\,14K) the emissivity 
of the cold component in the bimodal model has to be enhanced. 

It is noteworthy, however, that the 
observed low $\tau_{200}$/A$_{\rm V}$ values
(\lele{14K}{T}{16K}), can be well fitted by using $\epsilon=1$
for both the warm and cold components.

This result suggests that the presence of colder component with the same
grain properties as for the DISM is sufficient to explain the low
$\tau_{200}$/A$_{\rm V}$ in a region. The most plausible reason is that
radiative transfer in these regions can cause some parts to become colder
while our temperature measurement is biased towards the warmer regions.
However, a necessary condition is that the colder component is well mixed
and not resolved by ISOPHOT, otherwise the scatter diagrams would become
highly non-linear. Whether this can be achieved in a cloud model, needs
to be investigated.
%
%We note that enhancing the relative contribution of the small grain
%component in the size distribution could lower the emissivity, since for
%a given grainsize $a$, $Q_{abs}\propto~a$ at a given wavelength. If this
%is a likely mechanism then the regions with $\tau_{200}$/A$_{\rm V}$ lower than
%the DISM would have undergone some recent processing such that the size
%disribition is biased towards smaller grainsizes.

% These temperatures are very low, and are characteristic for the
% densest cores of the interstellar medium. For such cores to exist
% in an otherwise unobscured line of sight, the interstellar medium 
% has to be very clumpy, with dense, but relatively low mass 
% (and therefore small) cores....???

%\subsection{Separation by environment?}

%There is a general trend, that $\epsilon$ values close to, or below the
%DISM value occur typically in isolated clouds,  while high emissivity
%values can be found in clouds, which are part of larger complexes (e.g.
%Taurus, $\rho$\,Oph). One  exception is LDN\,1251 (G114.0+14.9,
%G114.3+14.7, G114.6+14.6),  which is apparently an isolated cloud, 
%however, it shows high $\epsilon$ values. Despite its location, it is
%well known, that this cloud belongs to the Cepheus-bubble (Kun \& Prusti
%1993, Kun 1998) therefore cannot be considered  as a real isolated
%cloud.} 

\subsection{Smaller grains}

An alternative explanation for low emissivity clouds with low
column density is an enhancement of the relative contribution 
of small grains, since for a grain size $a$,  
$Q_{abs}\,{\propto}\,a$ in the far-infrared.
If this is a likely mechanism then the regions 
with $\tau_{200}$/A$_{\rm V}$ lower than the DISM would have undergone 
some recent processing such that the relative amount of small
grains is larger than in the standard grain size distribution.
 We note that a fraction of the clouds were 
selected from the IRAS data on the basis of the high 100\,$\mu$m 
brightness. These clouds also exhibit a high 
brightness in the IRAS 12, 25, and 60\,$\mu$m 
bands and could indicate a bias towards clouds with a 
grain size distibution favouring the smaller sizes.
%%%%%%%%%%%%%%%%%%%%%%%%%%%%%%%%%%%%%%%%%%%%%%%

\section{Conclusions}

In this paper we have analysed the FIR emission properties in a large 
sample of interstellar clouds, observed with ISOPHOT.  We have derived
far-infrared emissivity relative to the visual extinction, \tauav{200},
for each region. The derived values of \tauav{200} represent the average
emissivity over a region typically in the order of 100\,arcmin$^2$.

The derived FIR emissivities \tauav{200} show an enhancement for the
coldest and densest regions where \lele{12\,K}{T$_{\rm d}$}{14\,K}, which is
most probably due to the growth of dust grains. We confirm a similar
trend found in earlier papers.
The enhancement of \tauav{200} for \lele{12\,K}{T$_{\rm d}$}{14\,K} is for the
majority of the regions less than 2. This is lower than previously
reported for some specific clouds. Our findings support models where the
enhancement in emissivity is attributed to ice mantle growth and the
presence of silicate agregrates on the spatial scales we investigated.

For \lele{14\,K}{T$_{\rm d}$}{17.5\,K}, we observe a majority of regions where
\tauav{200} is lower than that of the diffuse ISM. FIR emissivities
lower than that of the DISM may be explained by the common effect of a
constant emissivity and the presence of multiple dust temperatures along
the line of sight. To sufficiently fit the observed values the cold
temperatures (T$_{\rm c}$) had to be set to  $\sim$12K without significantly
altering the dust properties. Alternatively, these clouds in our sample
could be biased towards regions with a more significant small grain component in
their dust size distribution.

%%%%%%%%%%%%%%%%%%%%%%%%%%%%%%%%%%%%%%%%%%
\section{Acknowledgments}

We thank the referee Dr. Mika Juvela for detailed comments and 
suggestions which helped us to improve the manuscript. 
This paper is based on observations with ISO an ESA project with
 instruments funded by ESA member states (especially the PI countries:
 France, Germany, the Netherlands and the United Kingdom) and with
 participation of ISAS and NASA. 
The {\em ISOPHOT\/} data were  processed
 using PIA, a joint development by the ESA Astrophysics Division and
 the {\em ISOPHOT\/} consortium led by MPI f\"ur Astronomie, Heidelberg.
 Contributing Institutes are DIAS, RAL, AIP, MPIK, and MPIA.
 
This research has made use of the following catalogues/services:
\begin{itemize}
\item USNOFS Image and Catalogue Archive, operated by the United States 
   Naval Observatory, Flagstaff Station (http://www.nofs.navy.mil/data/fchpix/)
\item NASA/IPAC Infrared Science Archive, which is operated by the 
   Jet Propulsion 
   Laboratory, California Institute of Technology, under contract
   with the National Aeronautics and Space Administration (the 2MASS
   point source catalogue)
\item ISO Data Archive, European Space Astronomy Centre of the   
  European Space Agency (http://www.iso.esac.esa.int/ida/)   
\end{itemize}  

This research has been supported by the grants T34584 and K62304 
of the Hungarian Research Fund. 

%%%%%%%%%%%%%%%%%%%%%%%%%%%%%%%%%%%%%%%%%%%%%%%%%%%%%%%%%%%%%%%%%%

%%%%%%%%%%%%%%%%%%%%%%%%%%%%%%%%%%%%%%%%%%%%%%%%%%%%%%%%%%%%%%%%%%%%%%%
%% ---------------------  APPENDIX -------------------- 
%%%%%%%%%%%%%%%%%%%%%%%%%%%%%%%%%%%%%%%%%%%%%%%%%%%%%%%%%%%%%%%%%%%%%%%

\clearpage
\appendix
\section{Comparison of the ISOPHOT and COBE/DIRBE 
surface brightness photometric systems}

The calibrational accuracy of the ISOPHOT surface brightness 
photometric system 
is crucial for the interpretation of the final results. 
Due to the lack of proper extended standard objects on the sky, especially 
at far-infrared wavelengths, the only practical possibility to test
the accuracy is a comparison of ISOPHOT data with that
of COBE/DIRBE values. 
Here first we check the general relationship of the ISOPHOT and COBE/DIRBE
surface brightness systems via the comparison of P22 mini-map 
highly processed data product (HPDP) background values from the
ISO Data Archive.
In a second approach average surface brightness values of our target fields
of the present study, obtained from ISOHPOT observations is compared with those 
derived from COBE/DIRBE surface brightness data. 
%We assume in this whole Appendix, that for both ISOPHOT and COBE/DIRBE 
%measurements the true surface brightness values are in a linear relationship
%with the measured ones, for the whole surface brightness range investigated. 

\subsection{Description of the COBE/DIRBE background determination}

The monochromatic COBE/DIRBE surface brightness value at a specific 
wavelength, date and sky position is derived with the {\sl PredictDIRBE} 
tool, written in IDL\footnote{Research Systems Inc., Versions 5.x and 6.x} 
and developed in Konkoly Observatory, using some routines written at
Max-Planck-Institut f\"ur Astronomie, Heidelberg.  
The {\sl PredictDIRBE} routine uses the 
CGIS\footnote{http://lambda.gsfc.nasa.gov/product/cobe/cgis.cfm} software 
library, developed to help the analysis of COBE data.  
PredictDIRBE uses two main DIRBE data products, the {\sl DIRBE
sky and zodi atlases} (DSZA)
and the {\sl zodi-subtracted mission average maps} (ZSMA).   
The detailed description of these DIRBE data products can be found on the COBE/DIRBE
homepage of IPAC\footnote{http://lambda.gsfc.nasa.gov/product/cobe/dirbe\_overview.cfm}. 
The main steps of the COBE/DIRBE surface brightness determination
in the {\sl PredictDIRBE} routine are the following: 
%%%%%%%%%%%%%%%%%%%%%%%%%%%%%%%%%%%%%%%%%%%%%%%%%%%%%%%%%%%%%%%%%%%%%%%
\begin{itemize}
\item {\bf Extraction of DIRBE ZSMA (zodiacal light component removed) surface 
  brightness values for the 10 DIRBE photometric bands}. 
  The number of DIRBE pixels considered here as well as for the determination 
  of the zodiacal contribution depends on the extension of ISOPHOT maps. 
  However, the spatial sampling of DIRBE measurements at long wavelenghts
  is very fine relative to the physical resolution ($\sim$30\arcmin or worse).  
  The operations below are done for the median values of the DIRBE pixels taken. 
\item {\bf Colour correction of the cirrus component}. In this step the temperature
  of the cirrus component is derermined by fitting the 100, 140 and 240\,$\mu$m 
  ZSMA surface brightness values with a modified black-body spectral energy 
  distribution (SED), I$_{\nu}$\,$\propto$\,$\nu^{\beta}$B$_{\nu}\rm(T)$, where
   B$_{\nu}\rm(T)$ is the Planck-function at temperature T 
   and frequency $\nu$, and
   $\beta$ is the spectral index. The 140\,$\mu$m DIRBE band has a lower
   weight in the fitting process, due to its well-known noisy behaviour 
   compared to the other bands. 
   For our calculations the spectral index has been fixed to $\beta$\,=\,2. 
   For $\lambda$\,$<$\,100\,$\mu$m the SED is approximated by
   a function fitted by spline interpolation to the measured 
   log$(\lambda)$\,--\,log(I$_{\lambda}^{\rm DIRBE})$ values.  
   The colour correction is performed using the fitted SEDs, and
   the transmission curves of the DIRBE filters. The final, monochromatic surface 
   brightness values are reached by the repetition of this process, until the 
   convergence criterium is matched. We have to note, that the ZSMA surface brightness 
   contains at least two main components: the cirrus (or, in general Galactic
   interstellar matter) emission and the extragalactic background. 
   Although the two components have different SEDs, 
   the difference between colour corrections needed for the two different SEDs is small, 
   and in most cases the absolute level of the extragalactic
   background component is much below that of the cirrus.  Therefore here we do not consider
   these two components separately.     
\item {\bf Determination of the non-zodiacal DIRBE surface brightness at the ISOPHOT
  wavelength}. The monochromatic DIRBE surface brightness values are interpolated 
  to the nominal ISOPHOT filter wavelength. For $\lambda$\,$<$\,100\,$\mu$m this 
  is performed by the spline interpolation of the 
  log$(\lambda)$\,--\,log(I$_{\lambda}^{\rm DIRBE})$ values, while for 
  $\lambda$\,$\ge$\,100\,$\mu$m the previously mentioned 
  $\nu^{\beta}$B$_{\nu}\rm(T)$ is applied. 
\item {\bf The zodiacal component} is determined from the DIRBE zodi 
 atlases (DSZA products). 
  From the date of the observation the actual solar elongation angle of the 
  requested sky coordinate is calculated. In case no DIRBE observation was
  performed at that solar elongation and ecliptic latitude (due to the limited lifetime
  of the instrument), the actual solar elongation is "mirrored", so that the solar
  elongation with the same absolute value is taken on the other side of the Sun, and
  the zodiacal component is extracted at the corresponding dates and coordinates. 
  This latter position was in almost all cases observed by DIRBE 
  (for the details of COBE/DIRBE observations, see
  Hauser et al., 1998b). The zodiacal component is extracted for all the 10 DIRBE
  photometric bands. 
\item {\bf Colour correction of the zodiacal component} is performed assuming a  
  pure black-body SED around
  the peak of the zodiacal emission, and using spline interpolation for 
  notably shorter and longer wavelengths, applying otherwise the same iterative process as
  for the colour correction of the non-zodi component above. The 
  resulting monochromatic zodiacal surface brightness values are used to 
  interpolate to the requested ISOPHOT wavelength.
\item {\bf The final PredictDIRBE surface brightness} is the sum of the 
  monochromatic zodiacal and the non-zodiacal (cirrus + extragalactic background) components.     
\end{itemize}

%%%%%%%%%%%%%%%%%%%%%%%%%%%%%%%%%%%%%%

\subsection{Minimaps HPDPs}
\begin{figure*}
\includegraphics[width=8cm]{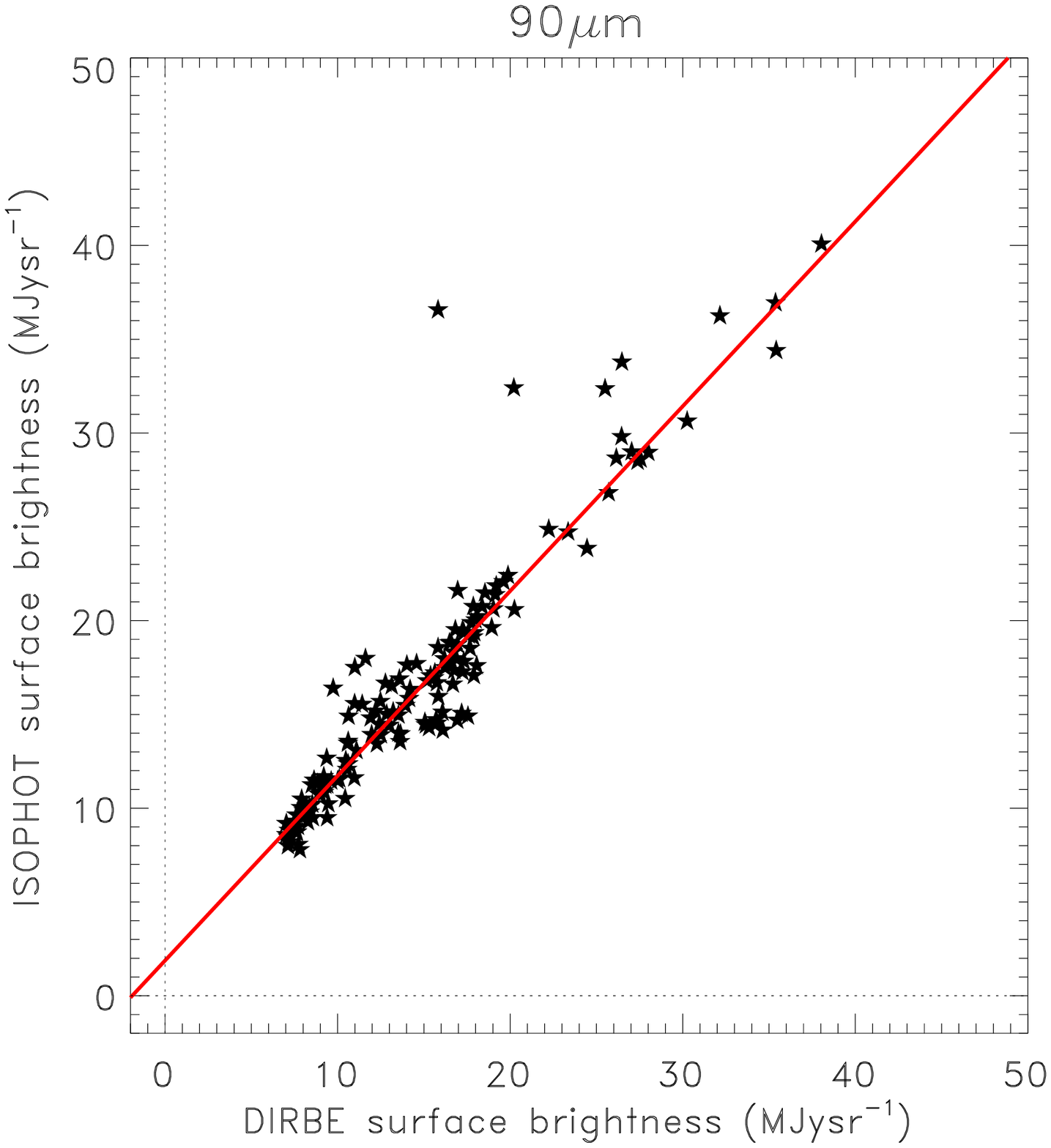}
\includegraphics[width=8cm]{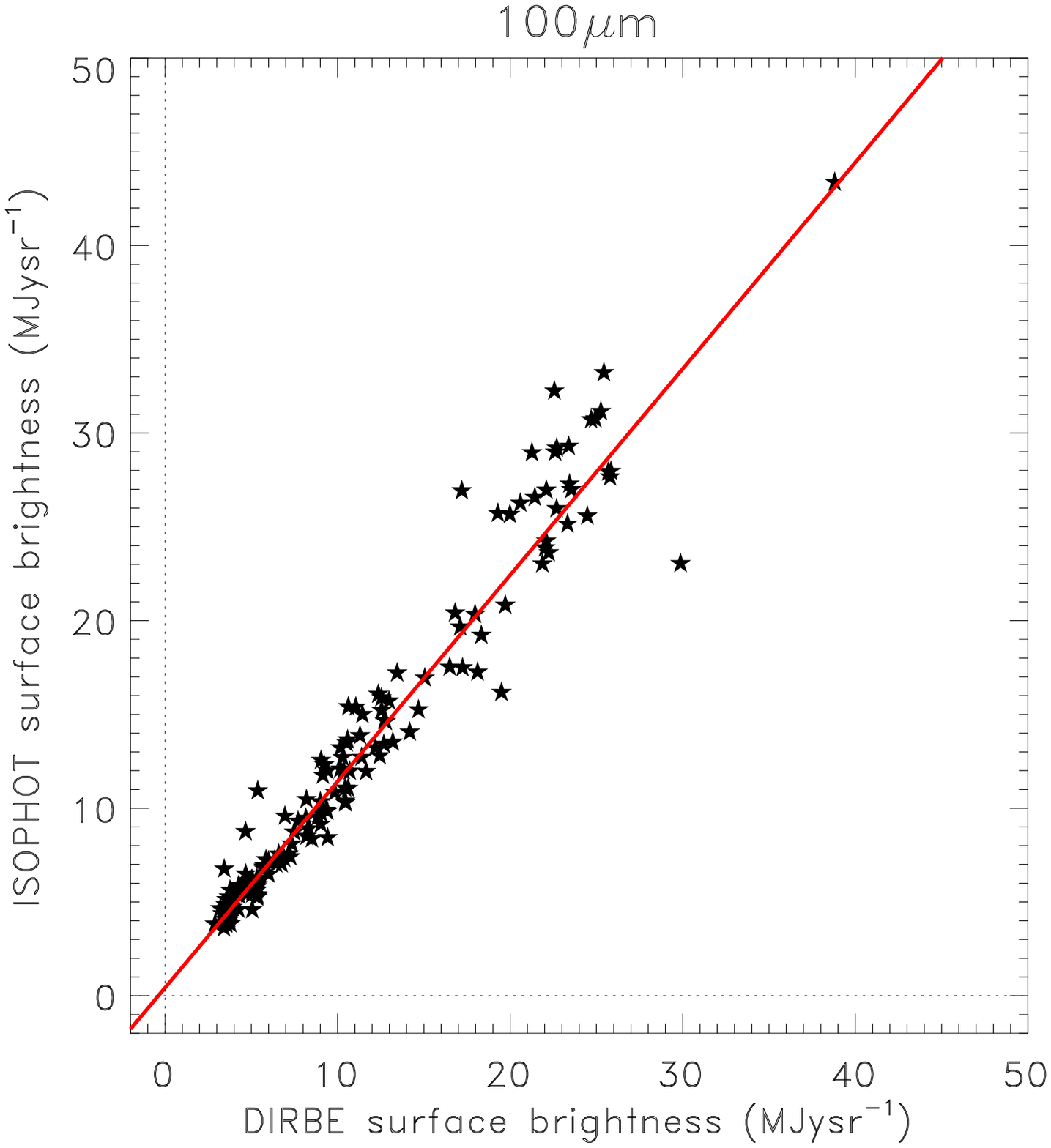}
\includegraphics[width=8cm]{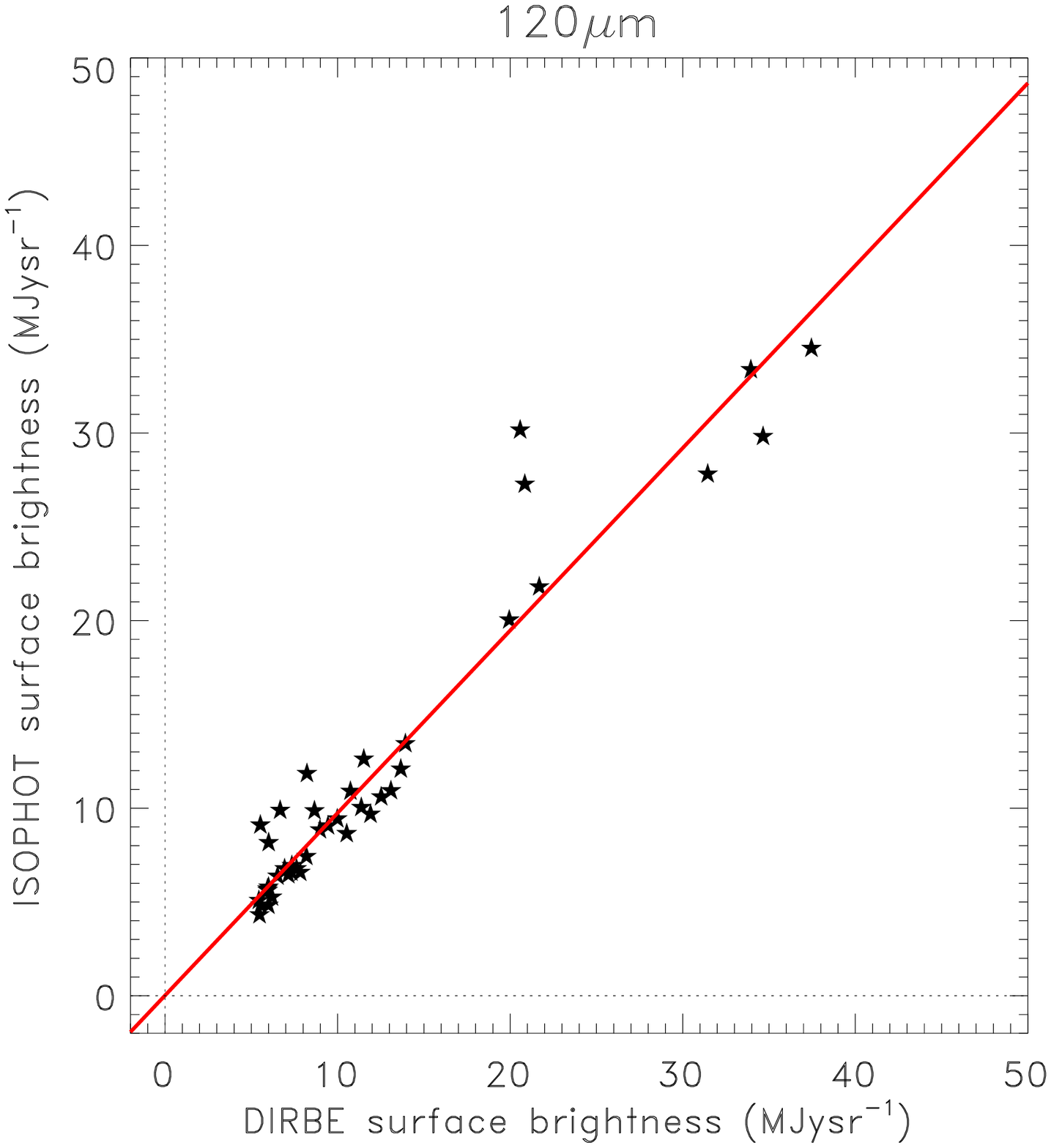}
\includegraphics[width=8cm]{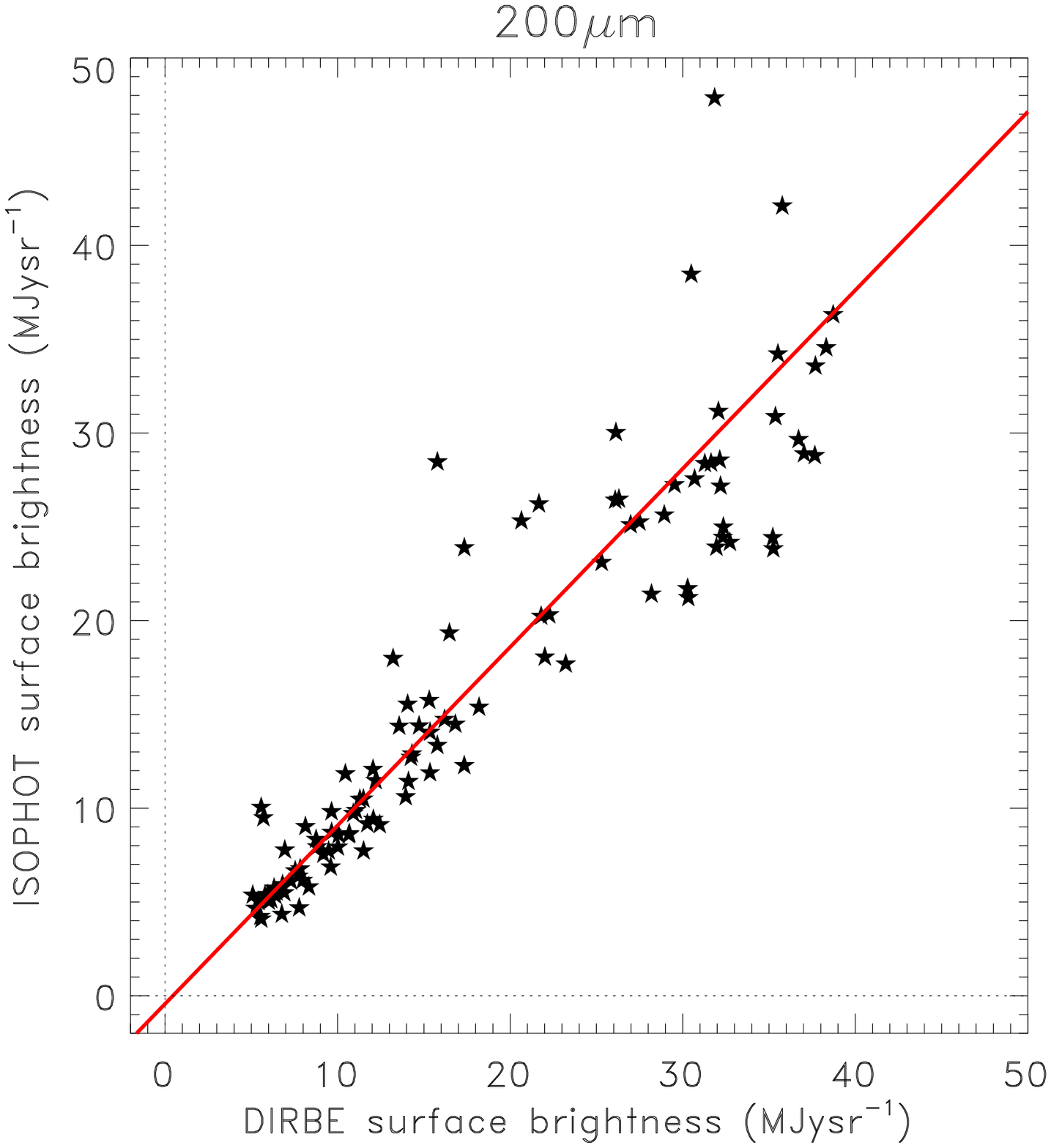}
\caption[]{Comparison of minimap background surface 
brightness values and background estimates based on 
COBE/DIRBE data at different wavelengths 
(a) 90\,$\mu$m, (b) 100\,$\mu$m, 
(c) 120\,$\mu$m and (d) 200\,$\mu$m. 
The solid lines are fitted linears, corresponding
to the values in Table~\ref{table-corr}.}
\label{fig:minimapcomp}
\end{figure*}

Mini-map mode was a 'submode' of the P22 astronomical observing 
template (AOT, see Laureijs et al. 2003) of ISOPHOT. The mode was
especially designed to obtain accurate fluxes of point/compact
sources. However, apart from the point source flux determination, 
the mini-maps provide accurate surface brightness values of the average 
background around the target, due to the redundant observation of the same
sky position by many ISOPHOT pixels. The mini-map backgrounds are assumed
to be homogeneous for the whole mini-map area. The typical spatial extension of 
a minimap is $\sim$4\arcmin~ for both the C100 and the C200 detector arrays.

The background values can be compared with the values obtained by the
{\sl PredictDIRBE} routine for the specific central coordinates of the source
(assuming a 30\arcmin~aperture) and for the ISOPHOT wavelengths. 
Mo\'or et al. (2003, 2004a, 2004b and 2005) performed a full re-evaluation 
of a large sample of ISOPHOT mini-maps, containing the observations of normal stars,
evolved stars, extragalactic and miscellaneous objects. 
The publicly available ISO Data 
Archive\footnote{http://www.iso.vilspa.esa.es/ida/index.html} 
contains the re-evaluated results of these
observation as Highly Processed Data Products (Salama et al., 2004). 

The results of the COBE/DIRBE and mini-map background comparisons  
are presented in Fig.~\ref{fig:minimapcomp} for
the four ISOPHOT filters used in our study (90, 100, 120 and 200\,$\mu$m). 
In Table~\ref{table-corr} we present the coefficients found for the correlation
between the DIRBE and ISOPHOT surface brightness photometric systems, assuming the 
linear relationship 
\begin{equation}
 \mathrm I_{\lambda}^{\rm ISOPHOT}\,=\,{\rm S}{\times}I_{\lambda}^{\rm DIRBE} + {\rm Offset} 
\label{eq-dirbe} 
\end{equation} 
where I$_{\lambda}^{\rm ISOPHOT}$ and I$_{\lambda}^{\rm DIRBE}$ are the 
surface brightness values at the wavelength $\lambda$, measured with 
ISOPHOT and DIRBE, respectively.
 
\begin{table}
\begin{tabular}{rrrrr}
\hline
$\lambda$ & S & Offset & R$_{1:1}$ & R$_{f}$ \\
($\mu$m)  &      & (MJy\,sr$^{-1}$) & (\%) & (\%) \\ 
\hline
90   & 0.98($\pm$0.02) & 1.87($\pm$0.13)   & 14.2 &  8.5 \\
100  & 1.09($\pm$0.03) & 0.43($\pm$0.13)   & 16.7 & 10.5 \\
120  & 0.97($\pm$0.06) & 0.02($\pm$0.28)   & 12.8 & 11.4 \\
200  & 0.95($\pm$0.03) & --0.44($\pm$0.24) & 14.9 & 10.8 \\
\hline
\end{tabular}
\caption[]{Parameters describing the relation between the DIRBE and 
ISOPHOT surface brightness photometric systems, derived from the mini-map 
HPDP sample, assuming a linear relationship. The columns of the table are:
(1) ISOPHOT filter nominal wavelength;
(2) fitted scaling factor (S), assuming Eq.~\ref{eq-dirbe};
(3) fitted offset, assuming Eq.~\ref{eq-dirbe}; 
(4) mean relative deviation of original ISOPHOT and DIRBE surface 
brightness values;
(5) mean relative deviation of ISOPHOT and DIRBE surface brightness
values after the correction with the fitted line;}
\label{table-corr}
\end{table}

The last two columns of Table~\ref{table-corr} contain
the mean relative deviations of orignal ISOPHOT and DIRBE surface brightness
values (column \#4) and the mean relative deviations of the original ISOPHOT 
and DIRBE surface brightness values after the correction with the fitted line
('residual scatter', column \#5). 
These values are representative for the general 
relative accuracy of the two photometric systems. 

The DIRBE surface brightness calibration (as well as that of ISOPHOT) 
contains intrinsic uncertainties, both in detector gain and offset. These 
uncertainties contribute significantly to the observed relative deviations 
listed in Table~\ref{table-corr}.  
For those DIRBE filters that are important for the comparison with the 
ISOPHOT surface brightness calibration in this paper, the typical gain uncertainties 
are 13.5\%, 10.6\% and 11.6\% for the 100, 140 and 240\,$\mu$m 
DIRBE bands, respectively  (see Arendt et al., 1998). 
These DIRBE gain uncertainties indicate the presence of an additional
uncertainty in the order of $\sim$10\% or below,
which should account for the uncertainties in the {\sl PredictDIRBE} estimate
calculation as well as in the intrinsic uncertainties of the ISOPHOT calibration. 

The typical offset uncertainties are
0.3, 0.6 and 0.4\,MJy\,sr$^{-1}$ for the 100, 140 and 240\,$\mu$m 
DIRBE bands per pixel, respectively  (Arendt et al., 1998). 
These are in the order of the offset values found for
the ISOPHOT--DIRBE comparison (see Table~\ref{table-corr}), except for the
90\,$\mu$m ISOPHOT filter, where this offset is significantly higher. 
However, the 90\,$\mu$m surface brightness values may also contain 
an emission contribution from a grain population different from the one
responsible for the $\lambda$\,$\ge$\,100\,$\mu$m part of the background SED. 
This effect is not accounted for in the PredictDIRBE routine. 

In the case of ISOPHOT, non-unity scaling factors may arise from
imperfect knowledge of the effective solid angles at different
wavelengths and/or from a flux-dependent non-linearity of the detector.
ISOPHOT solid angles have been re-assessed recently (\'Abrah\'am et al. 2005,
ISOPHOT internal calibration report) and the new values agree within 5\%
with the previous ones.
Non-linearity can occur due to calibration uncertainties in the
extrapolation to weak flux levels because the ISOPHOT calibration targets
were always much brighter than the ISM in the ISOPHOT beams. The
importance of the non-linearity can be estimated via comparison with the
COBE/DIRBE calibration.  According to the results presented in 
Table~\ref{table-corr} the relationships between the ISOPHOT and COBE/DIRBE surface 
brightness values are sufficiently linear.

%%%%%%%%%%%%%%%%%%%%%%%%%%%%%%%%%%%%%%%%%%%%%%%%%%%%%%%%%%%%%%%%%%%%%%%%%%%%%%%%%%%%%
\begin{figure}
\includegraphics[width=8cm]{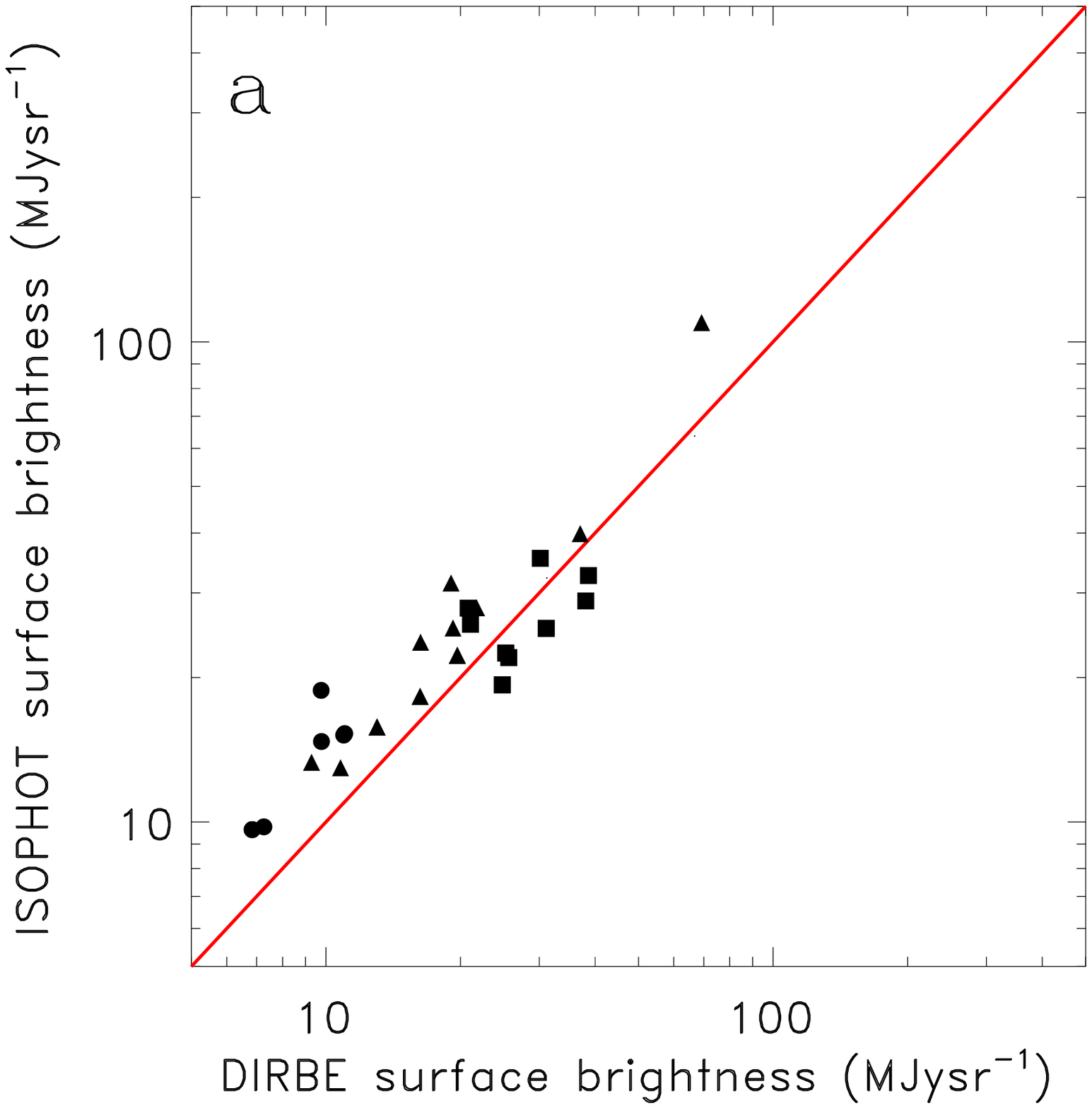}
\includegraphics[width=8cm]{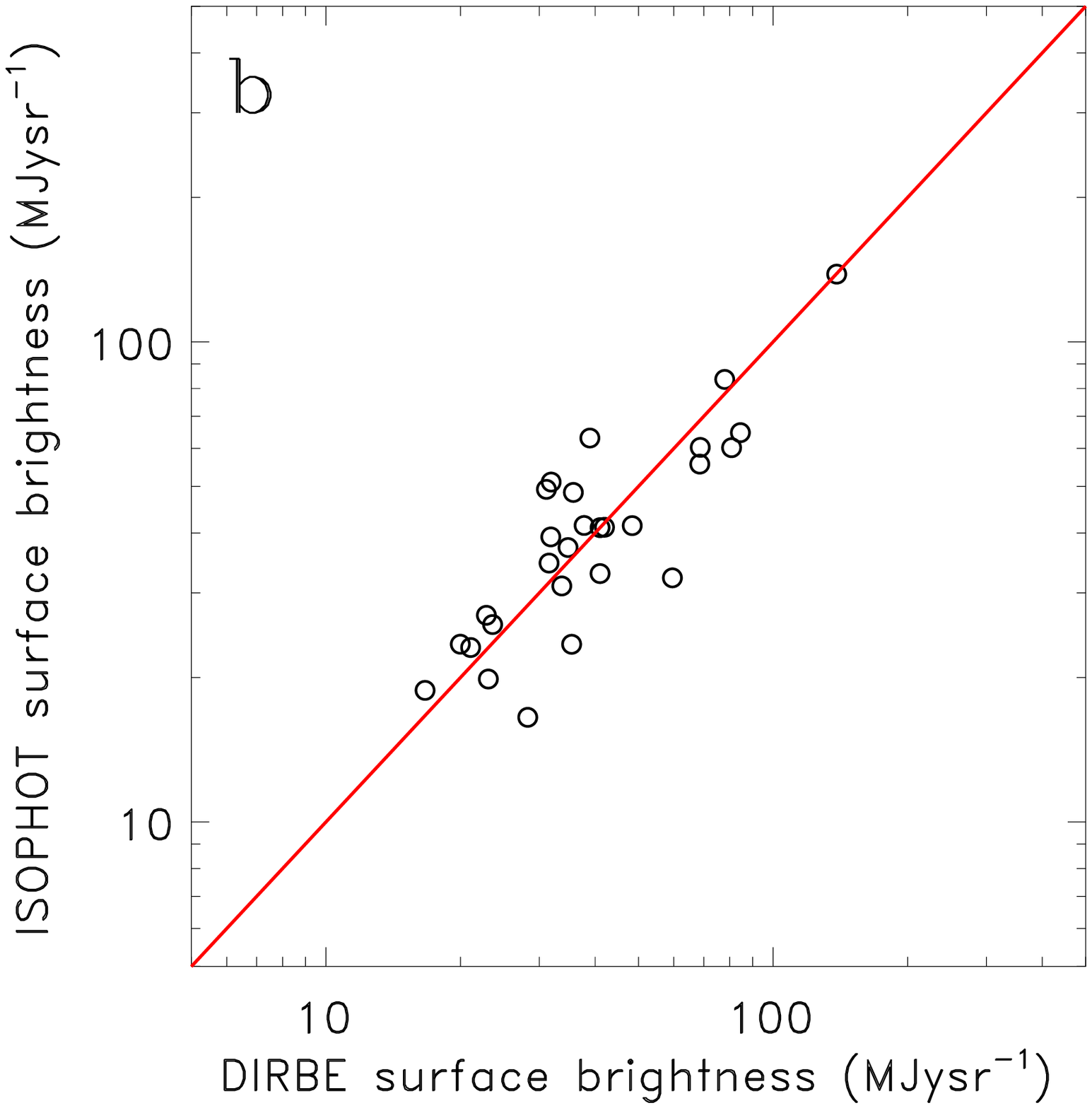}
\caption[]{Comparison of the average surface brightness of our
target fields and the DIRBE surface brightness derived for the same 
sky positions and observational dates, with the {\sl PredictDIRBE} routine. 
({\bf a}) Short wavelength filters (ISOPHOT 90, 100 and 120\,$\mu$m, 
marked by filled circles, triangles and squares, respectively); 
({\bf b}) ISOPHOT 200\,$\mu$m filter. The solid lines mark the 1:1 line.}
\label{fig-fieldcomp}
\end{figure}

\subsection{Fields of the present study}

The target fields of the present study were observed by the same AOT (P22) 
as the mini-maps, with some notable differences between the
two samples. First, the extension of the target field maps is in the 
5\arcmin...40\arcmin range, while the mini-maps are smaller than these. 
Second, the intensity of the target maps contains the contribution 
of point sources and also that of larger scale structures, while the 
point source emission was subtracted in the derivation of 
the mini-map backgrounds. 
     
The average surface brightness of our target fields were derived as simple 
average of all pixel values, for both the short and the long wavelength maps
of a specific field. {\sl PredictDIRBE} estimates of the DIRBE surface 
brightness were calculated for the central positions of the ISOPHOT maps. 
The observational date of the maps were also considered to account for the
annual variation of the zodiacal light component. The results are 
presented in Fig.~\ref{fig-fieldcomp} and in Table~\ref{table-dirbefield}.

%%%%%%%%%%%%%%%%%%%%%%%%%%%%%%%%%%%%%%%%%%%%%%%%%%%%%%%%%%%
\begin{table}
\begin{tabular}{lrrrrr}
\hline
field & $\lambda_{s}$/$\lambda_{l}$ & I$_{s}^{D}$ & I$_{s}^{PHT}$ 
 & I$_{l}^{D}$ & I$_{l}^{PHT}$ \\
 % &  & (MJy\,sr$^{-1}$)  & (MJy\,sr$^{-1}$) 
 % & (MJy\,sr$^{-1}$) & (MJy\,sr$^{-1}$) \\
\hline
G004.3+35.8 &   100 / 200 &     19.0 &  31.5 &  38.9 &  63.1 \\
G100.0+14.8 &    90 / 200 &     11.0 &  15.3 &  31.5 &  34.6 \\
G101.8+17.0 &   100 / 200 &     10.8 &  13.0 &  23.1 &  19.9 \\
G102.0+15.2 &    90 / 200 &      9.8 &  14.7 &  37.8 &  41.5 \\
G114.0+14.9 &   120 / 200 &     24.8 &  19.3 &  41.0 &  32.9 \\
G114.3+14.7 &   120 / 200 &     25.6 &  22.0 &  42.0 &  41.1 \\
G114.6+14.6 &   120 / 200 &     25.2 &  22.5 &  41.0 &  41.0 \\
G121.6+24.6 &    90 / 200 &      7.3 &   9.8 &  22.8 &  27.0 \\
G122.0+24.2 &    90 / 200 &      6.8 &   9.6 &  21.1 &  23.1 \\
G142.0+38.5 &   100 / 200 &      9.3 &  13.3 &  16.7 &  18.8 \\
G170.2--16.0 &  120 / 200 &     31.1 &  25.3 &  48.4 &  41.4 \\
G173.9--15.7 &  120 / 200 &     38.1 &  28.9 &  68.5 &  55.7 \\
G174.3--15.9 &  120 / 200 &     38.6 &  32.6 &  68.7 &  60.3 \\
G297.3--16.2 &  100 / 200 &     21.7 &  27.9 &  84.5 &  64.7 \\
G300.2--16.8 &  120 / 200 &      9.8 &  18.8 &  31.8 &  39.2 \\
G301.7--16.6 &   90 / 200 &     16.3 &  23.7 &  34.8 &  37.3 \\
G302.6--15.9 &  100 / 200 &     13.0 &  15.8 &  28.3 &  16.5 \\
G303.5--14.2 &  100 / 200 &     19.7 &  22.2 &  59.5 &  32.2 \\
G303.8--14.2 &  100 / 200 &     16.2 &  18.3 &  35.4 &  23.5 \\
G355.3+14.7  &  100 / 200 &     69.2 & 109.7 & 138.8 & 138.4 \\
G359.1+36.7  &  100 / 200 &     19.2 &  25.3 &  23.6 &  25.8 \\
G359.9--17.9 &  100 / 200 &     37.0 &  39.8 &  78.0 &  83.5 \\
\hline
G089.0--41.2$^*$ &   90 / 200 & 11.0 & 15.2 & 20.0 & 23.5 \\
G111.2+19.6$^*$  &  150 / 200 & 31.2 & 32.2 & 33.7 & 31.0 \\
G187.3--16.7$^*$ &  120 / 200 & 30.2 & 35.4 & 35.8 & 48.6 \\
G297.3--15.7$^*$ &  150 / 200 & 66.8 & 63.7 & 80.8 & 60.2 \\
G301.2--16.5$^*$ &  120 / 200 & 20.8 & 27.9 & 31.1 & 49.3 \\
G301.7--16.6$^*$ &  120 / 200 & 21.1 & 25.8 & 31.9 & 51.1 \\
\hline
\end{tabular}

\caption{
 The columns of the table are:
(1) the name of the field (central galactic coordinates);
(2) central wavelengths of the short/long ISOPHOT filters [$\mu$m/$\mu$m]; 
(3) short wavelength DIRBE surface brightness [MJy\,sr$^{-1}$];
(4) short wavelength ISOPHOT surface brightness [MJy\,sr$^{-1}$];
(5) long wavelength DIRBE surface brightness [MJy\,sr$^{-1}$];
(6) long wavelength ISOPHOT surface brightness [MJy\,sr$^{-1}$] }
\label{table-dirbefield}
\end{table}
%%%%%%%%%%%%%%%%%%%%%%%%%%%%%%%%%%%%%%%%%%%%%%%%%%%%%%%%%%%

\subsection{Summary}

The results of the DIRBE and ISOPHOT surface brightness comparisons can be 
summarized as follows: 
\begin{itemize} 

\item In general a very good linear relationship was found between the
ISOPHOT and DIRBE surface brightness photometric systems.

\item 
The scaling factors between the DIRBE and the ISOPHOT
surface brightness photometric systems are close to unity, within the 
uncertainties. Even largest difference (100\,$\mu$m) is
within 9\% to the DIRBE system. These scaling factors indeed have an impact 
on the derived dust temperatures and \iav{200} ratios. 
This effect is widely discussed for the typical deivations in the main text
(see Sect.~4.5). 

\item The offsets found between the DIRBE and ISOPHOT 
surface brightness calibration based on the mini-map database 
are usually small (see Table~\ref{table-corr}). 
The largest offset value ($\sim$1.9\,MJy\,sr$^{-1}$) is found for the ISOPHOT 
90\,$\mu$m filter.  However, in this paper we applied the method of slope fitting in
the derivation of dust temperature from surface brightness scatter plots 
and in the determination of the \iav{200} ratio, which is completely
insensitive for offsets. 

\item Not taking into account the systematic differences,  
the mean relative deviations between the ISOPHOT the COBE/DIRBE
surface brightness values are within $\sim$15\% for the four
investigated wavelengths.   

\item The comparison of the ISOPHOT vs. DIRBE relative deviations with the 
DIRBE intrinsic detector gain uncertainties and the ISOPHOT vs. DIRBE 
offsets with the DIRBE intrinsic offset uncertainties show, that the
uncertainties in the DIRBE surface brightness calibration have a siginificant 
impact on the ISOPHOT--DIRBE comparison. The DIRBE detector calibration 
uncertainties are at a similar level than other error sources (e.g. ISOPHOT
detector calibration uncertainties, DIRBE background estimate uncertainties
by the PredictDIRBE routine) and they may be dominant in some cases.    

\item A representative ISOPHOT calibration uncertainty can be estimated
by subtracting the DIRBE detector gain uncertainties (in average $\sim$12\%
for the 100, 140 and 240\,$\mu$m DIRBE bands) from the mean relative
deviations between the two photometric systems (in average $\sim$15\%, see
column \#4 in Table~A1). The resulting value which we adopt for both short and
long ISOPHOT wavelengths is \emph{9\%}.  
 
\item The DIRBE and ISOPHOT surface brightness values in the target fields show 
a behaviour very similar to that of the mini-map backgrounds. The general 
agreement between the ISOPHOT and DIRBE surface brightness values are very good. 
The relatively large deviations in some cases may be explained by the 
beamsize of DIRBE, which is large, even compared to the size of the 
ISOPHOT target maps. 
  
\end{itemize}

\end{document}